\documentclass{cernrep}
\usepackage{graphicx,here,epsfig,user}
\usepackage{color}
\definecolor{GreenYellow}   {cmyk}{0.15,0,0.69,0}
\definecolor{Yellow}        {cmyk}{0,0,1,0}
\definecolor{Goldenrod}     {cmyk}{0,0.10,0.84,0}
\definecolor{Dandelion}     {cmyk}{0,0.29,0.84,0}
\definecolor{Apricot}       {cmyk}{0,0.32,0.52,0}
\definecolor{Peach}         {cmyk}{0,0.50,0.70,0}
\definecolor{Melon}         {cmyk}{0,0.46,0.50,0}
\definecolor{YellowOrange}  {cmyk}{0,0.42,1,0}
\definecolor{Orange}        {cmyk}{0,0.61,0.87,0}
\definecolor{BurntOrange}   {cmyk}{0,0.51,1,0}
\definecolor{Bittersweet}   {cmyk}{0,0.75,1,0.24}
\definecolor{RedOrange}     {cmyk}{0,0.77,0.87,0}
\definecolor{Mahogany}      {cmyk}{0,0.85,0.87,0.35}
\definecolor{Maroon}        {cmyk}{0,0.87,0.68,0.32}
\definecolor{BrickRed}      {cmyk}{0,0.89,0.94,0.28}
\definecolor{Red}           {cmyk}{0,1,1,0}
\definecolor{OrangeRed}     {cmyk}{0,1,0.50,0}
\definecolor{RubineRed}     {cmyk}{0,1,0.13,0}
\definecolor{WildStrawberry}{cmyk}{0,0.96,0.39,0}
\definecolor{Salmon}        {cmyk}{0,0.53,0.38,0}
\definecolor{CarnationPink} {cmyk}{0,0.63,0,0}
\definecolor{Magenta}       {cmyk}{0,1,0,0}
\definecolor{VioletRed}     {cmyk}{0,0.81,0,0}
\definecolor{Rhodamine}     {cmyk}{0,0.82,0,0}
\definecolor{Mulberry}      {cmyk}{0.34,0.90,0,0.02}
\definecolor{RedViolet}     {cmyk}{0.07,0.90,0,0.34}
\definecolor{Fuchsia}       {cmyk}{0.47,0.91,0,0.08}
\definecolor{Lavender}      {cmyk}{0,0.48,0,0}
\definecolor{Thistle}       {cmyk}{0.12,0.59,0,0}
\definecolor{Orchid}        {cmyk}{0.32,0.64,0,0}
\definecolor{DarkOrchid}    {cmyk}{0.40,0.80,0.20,0}
\definecolor{Purple}        {cmyk}{0.45,0.86,0,0}
\definecolor{Plum}          {cmyk}{0.50,1,0,0}
\definecolor{Violet}        {cmyk}{0.79,0.88,0,0}
\definecolor{RoyalPurple}   {cmyk}{0.75,0.90,0,0}
\definecolor{BlueViolet}    {cmyk}{0.86,0.91,0,0.04}
\definecolor{Periwinkle}    {cmyk}{0.57,0.55,0,0}
\definecolor{CadetBlue}     {cmyk}{0.62,0.57,0.23,0}
\definecolor{CornflowerBlue}{cmyk}{0.65,0.13,0,0}
\definecolor{MidnightBlue}  {cmyk}{0.98,0.13,0,0.43}
\definecolor{NavyBlue}      {cmyk}{0.94,0.54,0,0}
\definecolor{RoyalBlue}     {cmyk}{1,0.50,0,0}
\definecolor{Blue}          {cmyk}{1,1,0,0}
\definecolor{Cerulean}      {cmyk}{0.94,0.11,0,0}
\definecolor{Cyan}          {cmyk}{1,0,0,0}
\definecolor{ProcessBlue}   {cmyk}{0.96,0,0,0}
\definecolor{SkyBlue}       {cmyk}{0.62,0,0.12,0}
\definecolor{Turquoise}     {cmyk}{0.85,0,0.20,0}
\definecolor{TealBlue}      {cmyk}{0.86,0,0.34,0.02}
\definecolor{Aquamarine}    {cmyk}{0.82,0,0.30,0}
\definecolor{BlueGreen}     {cmyk}{0.85,0,0.33,0}
\definecolor{Emerald}       {cmyk}{1,0,0.50,0}
\definecolor{JungleGreen}   {cmyk}{0.99,0,0.52,0}
\definecolor{SeaGreen}      {cmyk}{0.69,0,0.50,0}
\definecolor{Green}         {cmyk}{1,0,1,0}
\definecolor{ForestGreen}   {cmyk}{0.91,0,0.88,0.12}
\definecolor{PineGreen}     {cmyk}{0.92,0,0.59,0.25}
\definecolor{LimeGreen}     {cmyk}{0.50,0,1,0}
\definecolor{YellowGreen}   {cmyk}{0.44,0,0.74,0}
\definecolor{SpringGreen}   {cmyk}{0.26,0,0.76,0}
\definecolor{OliveGreen}    {cmyk}{0.64,0,0.95,0.40}
\definecolor{RawSienna}     {cmyk}{0,0.72,1,0.45}
\definecolor{Sepia}         {cmyk}{0,0.83,1,0.70}
\definecolor{Brown}         {cmyk}{0,0.81,1,0.60}
\definecolor{Tan}           {cmyk}{0.14,0.42,0.56,0}
\definecolor{Gray}          {cmyk}{0,0,0,0.50}
\definecolor{Black}         {cmyk}{0,0,0,1}
\definecolor{White}         {cmyk}{0,0,0,0}

\usepackage{axodraw}
\usepackage{latexsym}
\usepackage{amssymb}

\def\mm{\mbox{$\mu^+\mu^-$}}

\def\gev   {{\rm GeV}}
\def\gevc  {{\rm GeV}}

\def\pb    {{\rm pb}}

\begin{document}

\title{Physics Opportunities at \mbf \mm\ Higgs Factories \footnote{\,\,Report 
       of the Higgs factory working group of the ECFA-CERN study on 
       Neutrino Factory \& Muon Storage Rings at CERN.}
       \vspace{-2cm}\flushright{\mbox{\rm\normalsize CERN-TH/2002-028}}}
 
\author{ 
C.~Bl\"ochinger$^{\,a}$, 
M. Carena$^{\,b}$, 
J. Ellis$^{\, c}$, 
H.~Fraas$^{\,a}$, 
F.~Franke$^{\,a}$, 
D.~Garcia$^{\,c}$, 
S.~Heinemeyer$^{\,d}$, 
S.~Kraml$^{\,c,e}$, 
G.~Moortgat-Pick$^{\,f}$, 
W.~Murray$^{\,g}$, 
F.~von~der~Pahlen$^{\,a}$, 
A.~Pilaftsis$^{\, a,h}$, 
C.E.M.~Wagner$^{\,i,j}$
and G.~Weiglein$^{\,c,k}$
}

\institute{ 
$a$ Inst. f. Theoretische Physik und Astrophysik, Univ. W\"urzburg, 
    D-97074 W\"urzburg, Germany\\[0.1cm]
$b$ Fermilab, P.O. Box 500, Batavia IL 60510, U.S.A.\\[0.1cm]
$c$ Theory Division, CERN, CH-1211 Geneva 23, Switzerland\\[0.1cm]
$d$ HET, Physics Department, Brookhaven Natl.\ Lab., 
    Upton, NY 11973 USA\\[0.1cm]
$e$ Inst. f. Hochenergiephysik, \"Osterr. Akademie d. Wissenschaften, 
    A-1050 Vienna, Austria\\[0.1cm]
$f$ DESY, Deutsches Elektronen-Synchrotron, D-22603 Hamburg, Germany\\[0.1cm]
$g$ RAL, Chilton, Didcot, Oxon., OX11 0QX, UK\\[0.1cm]
$h$ Department of Physics and Astronomy, Univ. of Manchester,
    Manchester M13~9PL, U.K.\\[0.1cm]
$i$ High Energy Physics Division, Argonne National Lab., 
    Argonne IL 60439, U.S.A.\\[0.1cm]
$j$ Enrico Fermi Institute, University of Chicago, 5640 Ellis Ave., 
    Chicago IL 60637, U.S.A.\\[0.1cm]
$k$ Institute for Particle Physics Phenomenology, 
    University of Durham, Durham DH1~3LR, UK
}
 
\maketitle 
 
\begin{abstract}

We update theoretical studies of the physics opportunities presented by
\mm\ Higgs factories. Interesting measurements of the Standard Model Higgs
decays into ${\bar b} b$, $\tau^+ \tau^-$ and $W W^*$ may be possible if
the Higgs mass is less than about $160~\gevc$, 
as preferred by the precision electroweak data, 
the mass range being extended by varying
appropriately the beam energy resolution. A suitable value of the beam
energy resolution would also enable the uncertainty in the $b$-quark mass
to be minimized, facilitating measurements of parameters in the MSSM 
at such a first \mm\ Higgs factory. These measurements would be sensitive
to radiative corrections to the Higgs-fermion-antifermion decay vertices,
which may violate CP. Radiative corrections in the MSSM may also induce CP
violation in Higgs-mass mixing, which can be probed via various
asymmetries measurable using polarized \mm\ beams. In addition,
Higgs-chargino couplings may be probed at a second \mm\ Higgs factory.

\end{abstract}


\section{Introduction}

Muon colliders produce Higgs bosons directly via \mm\ annihilation in the
$s$-channel, unaccompanied by spectator particles.  
If the electroweak symmetry is broken via the Higgs mechanism, 
hadron machines, such as the Tevatron collider \cite{TevHiggs} and the 
LHC \cite{LHCHiggs}, will presumably discover at least one Higgs boson, 
but in an experimental environment contaminated by important backgrounds 
and accompanied by many other particles. 
An $e^+ e^-$ linear collider (LC)~\cite{tesla,orangebook,jlc} 
would complement the hadron colliders by
providing precise studies of the Higgs boson in a clean environment.
However, the dominant production mechanisms create Higgs bosons in
association with other particles, such as a $Z^0$\,, two neutrinos or an
$e^+ e^-$ pair. Moreover, the peak cross section for a \mm\ collider to
produce a Higgs of $115~\gevc$ is around 60~pb, which can be compared
with around 0.14~pb for an $e^+e^-$ collider operating at 350~GeV.

The potential of \mm\ colliders for investigations of the Higgs
system is very exciting, and has been the subject of much work, see,
e.g.,~\cite{USworkshops,cern99,Barger:2001mi}. 
However, if the study of an $s$-channel resonance is
to be pursued experimentally, the event rate must be sufficiently large.
In the case of a Standard Model (SM) Higgs boson $H$, this means that the
mass must be somewhat less than twice $M_W$, otherwise the large width
reduces the peak cross section. This condition need not apply to more
complicated Higgs systems, for instance the heavier neutral Higgses of
supersymmetry.

\begin{figure}[htb!]
\begin{center}
\epsfig{figure=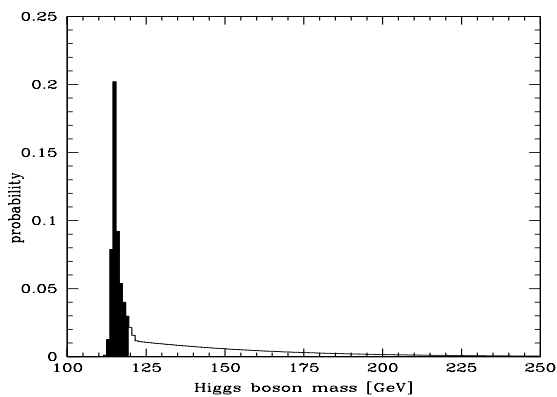,width=7.5cm,height=5.4cm}
\end{center}  
\caption[]{Probability distribution for the mass of the SM Higgs boson,
estimated~\cite{erler} by combining the available indirect information
with the LEP direct lower limit~\cite{LEPtb}. 
The shaded region represents 50\% of the probability distribution.
}
\label{fig:erler}
\end{figure}

Since a \mm\ collider is able to work near optimally over only a limited
range of centre-of-mass energies, knowledge of the Higgs mass is crucial
in designing such a machine.  A combined fit to precision electroweak
observables yields an indirect estimate for the SM Higgs boson mass of
\begin{equation}
  m_H \, = \, 88^{\,+53}_{\,-35}~\:\gevc \,.
\label{pewf}
\end{equation}
with a one-sided 95\% confidence-level upper limit of
$196~\gevc$~\cite{SM-fits}, including theoretical uncertainties. These
numbers are increased by about $20~\gevc$ if one uses the
estimate~\cite{ADM} of the effective value of $\alpha_{em}$ at the $Z^0$
peak. The range (\ref{pewf}) should be compared with the lower limit from
direct searches of $114.1~\gevc$~\cite{LEPtb}, and suggests that the most
probable value for the Higgs mass is not much greater than this lower
limit~\cite{erler}, as seen in Fig.~\ref{fig:erler}.
The analysis leading to Fig.~\ref{fig:erler} is valid within 
the Standard Model, or any new physics extension of it in which
the new physics effects decouple from the precision electroweak 
observables, as occurs for example in minimal supersymmetric 
extensions  of the Standard  Model, when all supersymmetric 
particle masses are  above the  weak scale.

In fact, the 2000 run of the LEP collider yielded a 2.1~$\sigma$ 
excess in the search for the SM Higgs boson, with a preferred 
mass of \cite{LEPtb} 
\begin{equation}
   m_H \, = \, 115.6^{\,+1.4}_{\,-1.1}~\:\gev\,.
\label{heroes}
\end{equation}
The excess seen is consistent with the expectations from
such a signal: the most significant candidate events have been seen in the 
$H \to \bar bb$ decay mode, with $Z^0 \to \bar qq$,
and the production cross section is quite compatible with that expected
for a SM Higgs boson.
The mass (\ref{heroes}) is highly consistent with the range
(\ref{pewf}). Moreover, both are also highly compatible with the minimal 
supersymmetric extension of the Standard Model (MSSM), which predicts the 
existence of a light Higgs boson weighing less than about 
$130~\gevc$~\cite{mhiggscorr}.
If the observation (\ref{heroes}) were to be confirmed, it would provide
an excellent opportunity for a \mm\ collider Higgs factory. As was outlined
in~\cite{cern99}, the measurement of the $H \to {\bar b} b$ decay mode for
a mass around $100~\gevc$ suffers from excessive background if $m_H$ is
close to the $Z$ peak, and from the rapidly-increasing Higgs width, and
therefore reduced on-peak cross section, as $m_H$ increases toward the
$W^+W^-$ threshold.  The optimal Higgs mass identified in~\cite{cern99}
was in fact $115~\gevc$.

In the coming years, first the Tevatron collider~\cite{TevHiggs} and
subsequently the LHC~\cite{LHCHiggs} will have opportunities to discover
the SM Higgs boson. 
In the case of the (constrained) MSSM, it has been shown that the
prospects for the lightest Higgs boson are nearly as good as for the
SM~\cite{ehow}. 
One may expect to measure the mass of the SM or MSSM Higgs boson at the
hadron colliders with a
precision better than $1~\gevc$.  Detailed follow-up measurements would
then be possible with an $e^+ e^-$ linear
collider~\cite{tesla,orangebook,jlc}.  
Unfortunately, the existence of a Higgs boson weighing $115~\gevc$ will
probably not be clarified by the Tevatron collider or the LHC for several
more years. 

On the other hand, much work is still required before the feasibility of a
\mm\ collider can be demonstrated. We recall that the muon collection and
storage facility foreseen for a \mm\ collider has many parameters in
common with those required for a neutrino factory \cite{cern99}, whose 
storage ring
requires $10^{14}$ muons per second to be injected with a preferred energy
of 50~GeV. This energy is close to that required for a
first-generation \mm\ collider. However, a \mm\ collider would need about
an order of magnitude more muons than are foreseen in the neutrino
factory, and it is not yet clear what combination of higher-efficiency
beam preparation and increased proton power will be the most effective way
to achieve this. Moreover, the normalised emittance envisaged for a
neutrino factory is 1.67 mm.rad, whereas 0.2 mm.rad is anticipated in \mm\
collider designs~\cite{ref:status_report}. Thus, considerably more beam
cooling would be required for a \mm\ collider. We recall also that the
bunch structure foreseen for a neutrino factory, namely a train of 140
bunches injected at 75~Hz, would need to be modified. The luminosity of a
collider scales with the square of the bunch current squared multiplied by
the repetition rate. To convert the neutrino factory into a muon collider,
the basic repetition rate of 75~Hz is quite suitable, but one requires
just one bunch in each cycle. If this can be done, and a six-dimensional
emittance of $1.7 \times 10^{-10}~(\pi m)^3$ can be
achieved~\cite{ref:status_report}, a luminosity of
$10^{31}$~cm$^{-2}$\,s$^{-1}$ may be achieved, colliding beams with an
energy spread of 0.01\%.

In this report,
we revisit first the physics prospects for \mm\ collider SM Higgs
factories, examining in particular two effects that were overlooked
in~\cite{cern99}. One is the $WW^*$ decay mode, which is rather clean and
has a branching ratio of at least 8\% in the SM. The other is the effect
of the beam energy spread, for which we consider values larger than the
0.003\% considered previously. In this way, the range of SM Higgs masses
for which useful measurements of the cross sections can be made extends up
to about 160~GeV.

We recall that there is a richer Higgs sector in the MSSM, including 
three neutral Higgs bosons $h$, $H$ and $A$, where the first two 
have scalar couplings in the CP-conserving limit, and the latter 
pseudoscalar couplings. 
As was also discussed in~\cite{cern99} there, 
are excellent prospects for a \mm\ collider tuned to the similar masses of
the heavier neutral Higgs bosons $H, A$. If they weigh several hundred
$\gevc$ or more, these might be difficult to observe and study at the
LHC or a linear $e^+ e^-$ collider. A Higgs boson weighing as little as
$115~\gevc$ is not only {\it consistent} with supersymmetry, but even
seems to {\it require} something very like it, if the effective Higgs
potential is not to become unstable at a relatively low energy
scale~\cite{ER}. Thus, confirmation of the hint (\ref{heroes}) would also
be a strong encouragement to envisage a second \mm\ Higgs factory, even if
the $H$ and $A$ have not been observed directly.
In this context we study the influence of supersymmetric radiative 
corrections on the peak cross sections and branching ratios of $h,H,A$ 
compared to a SM Higgs boson. 

Both the first \mm $h$ factory (FMC) and the second \mm $(H,A)$ factory 
(SMC) will provide unique opportunities to study CP violation in the Higgs 
sector of the MSSM~\cite{cern99}. 
There have recently been improved studies of this
possibility~\cite{CEPW}, in the light of which we revisit here the
prospects for measuring various CP-violating observables at \mm\ colliders.
Finally, 
we also discuss the prospects for measuring the $H, A$ couplings to
charginos at such a second \mm $(H,A)$ factory.

\section{CP-Conserving Studies}

\setlength{\unitlength}{1mm}
 
\subsection{\mbf The $\mumu\to H\to X$ Cross Section}

The effective cross section for Higgs production at $\rts \sim m_H$ is 
obtained by
convoluting the standard $s$-channel Breit-Wigner resonance with the beam 
energy distribution, which we model as a
Gaussian distribution with width $\sigrts$. At $\rts=m_H$, initial-state 
radiation (ISR) effects can
be approximated by a constant reduction factor $\eta$, where
$\eta$ is a function of the various parameters, $\alpha$,
$m_H$, $m_\mu$, \ldots, that we do not discuss here. In the limit 
$\Gamma\ll m_H$, quite a
compact expression can be derived for the peak cross section:
\begin{equation}
  \s_{\rm peak} = \s(\rts=m_H) = 
  {4\pi\, B(H\to\mumu)\, B(H\to X) \over m_H^2}\: 
  \eta \, \pi^{1/2}\, A\, e^{A^2} (1-{\rm Erf}(A))
  \, ,
\label{peak-xs}
\end{equation}
where
\begin{equation}
  A = \frac{1}{2\sqrt{2}}\, \frac{\Gamma}{\sigrts}
  \qquad {\rm and} \qquad
  {\rm Erf}(x) = \frac{2}{\sqrt{\pi}}\, \int_0^x{e^{-t^2}dt} \,.
\label{a-defn}
\end{equation}

\noi
The peak cross section depends critically on the beam-energy spread  
$\sigrts$ compared to the resonance width $\Gamma$. 
There are two important limits:
\begin{eqnarray}
  \sigrts\,\ll\,\Gamma &\Rightarrow& 
    \sigma_{\rm peak} = {4\pi\,\eta\, B(H\to\mumu)\, B(H\to X) \over m_H^2}\,,
    \label{eq:sigapprox1}
    \\[1mm]
  \sigrts\,\gg\,\Gamma &\Rightarrow& 
    \sigma_{\rm peak} = {\sqrt{2\pi^3}\,\eta\, 
    \Gamma(H\to\mumu)\, B(H\to X) \over m_H^2\,\sigrts}\,.
    \label{eq:sigapprox2}
\end{eqnarray}

\begin{figure}[h!]
\begin{center}
\begin{picture}(148,58)
\put(0,0){\mbox{ 
  \epsfig{file=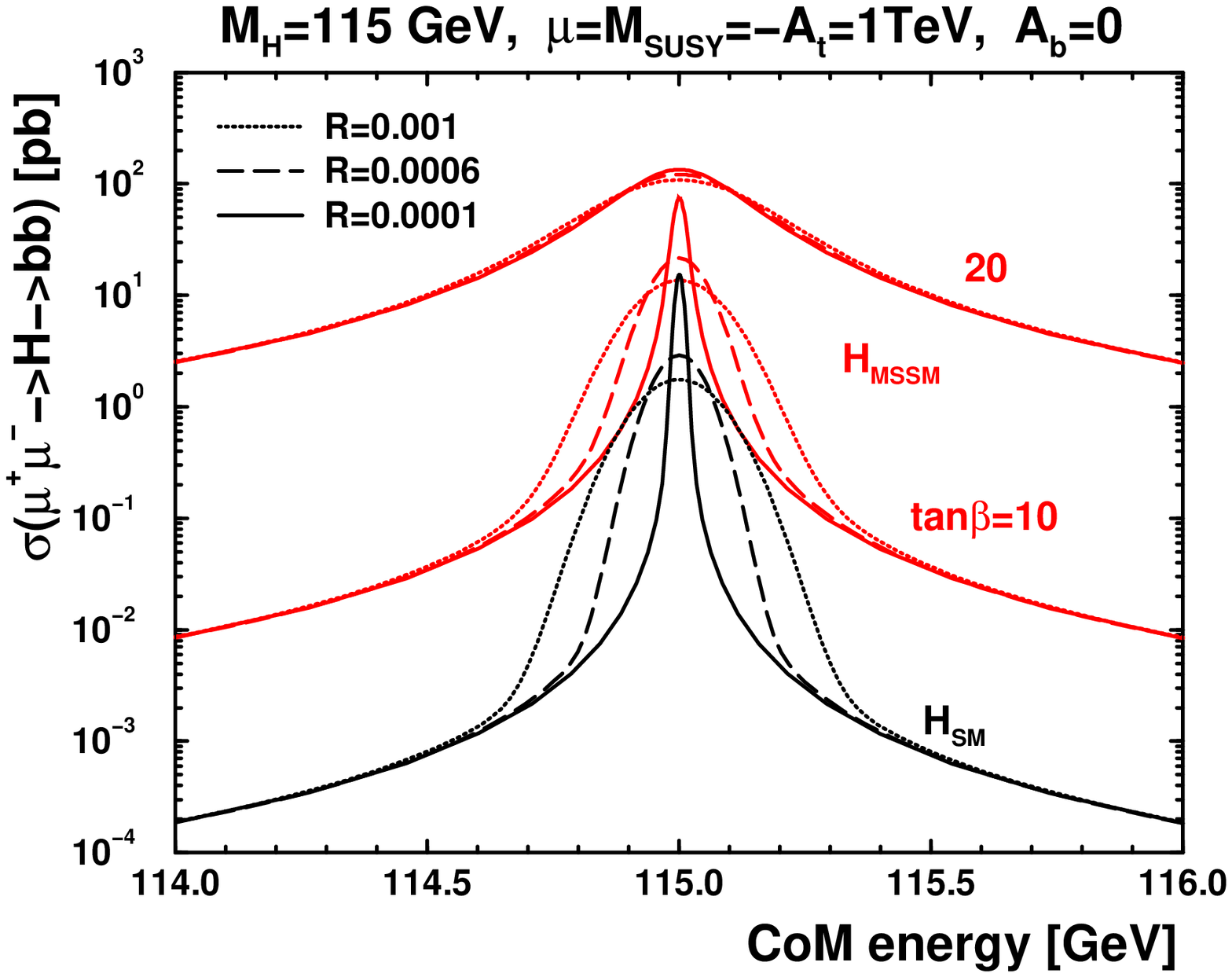,height=56mm} \hspace{8mm} 
  \epsfig{file=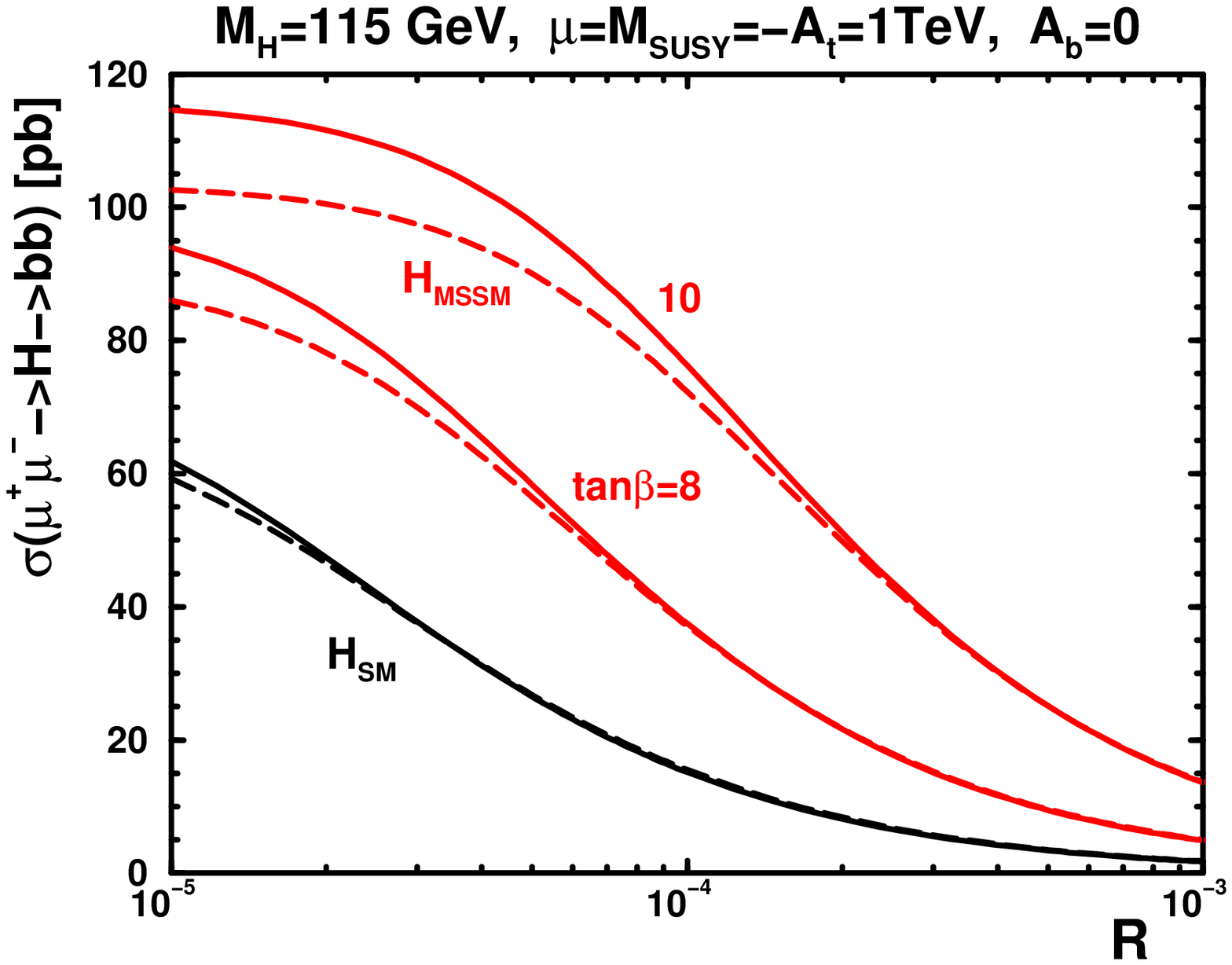,height=56mm} }}
\put(1,54){\mbox{\bf a)}}
\put(79,54){\mbox{\bf b)}}
\end{picture}
\end{center}
\vspace*{-5mm}
\caption{(a) Cross sections for $\mumu\to H\to\bb$ as functions of
  $\sqrt{s}$  for SM and MSSM Higgs bosons, and
  (b) $R$ dependences of the peak cross sections, for
  $\omb(\omb)=4.15$~GeV (solid lines) and $\omb(\omb)=4.45$~GeV
  (dashed lines).}
\label{fig:sigpeak}
\end{figure}

\noi
Figure~\ref{fig:sigpeak}(a) shows the $\sqrt{s}$ dependence of 
$\s(\mumu\to H\to\bb)$ for a SM Higgs boson $H_\mathrm{SM}$ weighing 
$115~\gevc$, compared with that
the lightest MSSM Higgs boson, denoted here by $H_\mathrm{MSSM}$,  
for various values of the beam-energy resolution 
$R\equiv\sqrt{2}\,\sigrts/\sqrt{s}$. 
ISR is neglected. 
The peak cross section is plotted 
as a function of $R$ in Fig.~\ref{fig:sigpeak}\,(b). 
As can be seen, $\s_{\rm peak}$ reaches a plateau for $R\ll\Gamma_{bb}/m_H$,
in accordance with (\ref{eq:sigapprox1}), (\ref{eq:sigapprox2}). 
Note also that the resonance is washed out in the limit 
$\Gamma/\sigrts\to 0$.

As has been discussed previously, not only is the beam energy spread at a 
\mm\ collider potentially very small, but also the energy can be 
calibrated 
very accurately using the decays of polarized muons in the circulating 
beams.
The very fine energy resolution and precision in $\sqrt{s}$ expected 
at the \mm\ collider would allow the properties of the 
Higgs boson(s) to be determined with outstanding accuracy. One
expects, for instance, to be able to measure the mass and width of a
light ($m_H<2\,M_W$) Higgs boson to fractions of an MeV.
If $\sigrts\lsim \Gamma$, the best procedure is to simply scan the
resonance, as was studied in detail 
in~\cite{bbgh96,USworkshops,cern99,Barger:2001mi}. 
For a very narrow resonance, e.g., for a light SM Higgs boson,
it may, however, be that $\sigrts\lsim \Gamma$ can only be achieved with 
substantial loss of luminosity. In this case it is of advantage to 
operate the collider at $\rts=m_H$ and two different beam energy 
resolutions $\sigma_{\!E}^{\rm min}\ll\Gamma$ and 
$\sigma_{\!E}^{\rm max}\gg\Gamma$ \cite{casal}. 
One can then determine the width of the resonance from the ratio of 
the peak cross sections:
\begin{equation}
  \s_{\rm peak}(\sigma_{\!E}^{\rm min}) / 
  \s_{\rm peak}(\sigma_{\!E}^{\rm max}) 
  = [2\sqrt{2}\sigma_{\!E}^{\rm min}]/[\sqrt{\pi}\,\Gamma\,] \, .
  \label{xsectionratio}
\end{equation}

\begin{figure}[htb!]
\begin{center}
\epsfig{figure=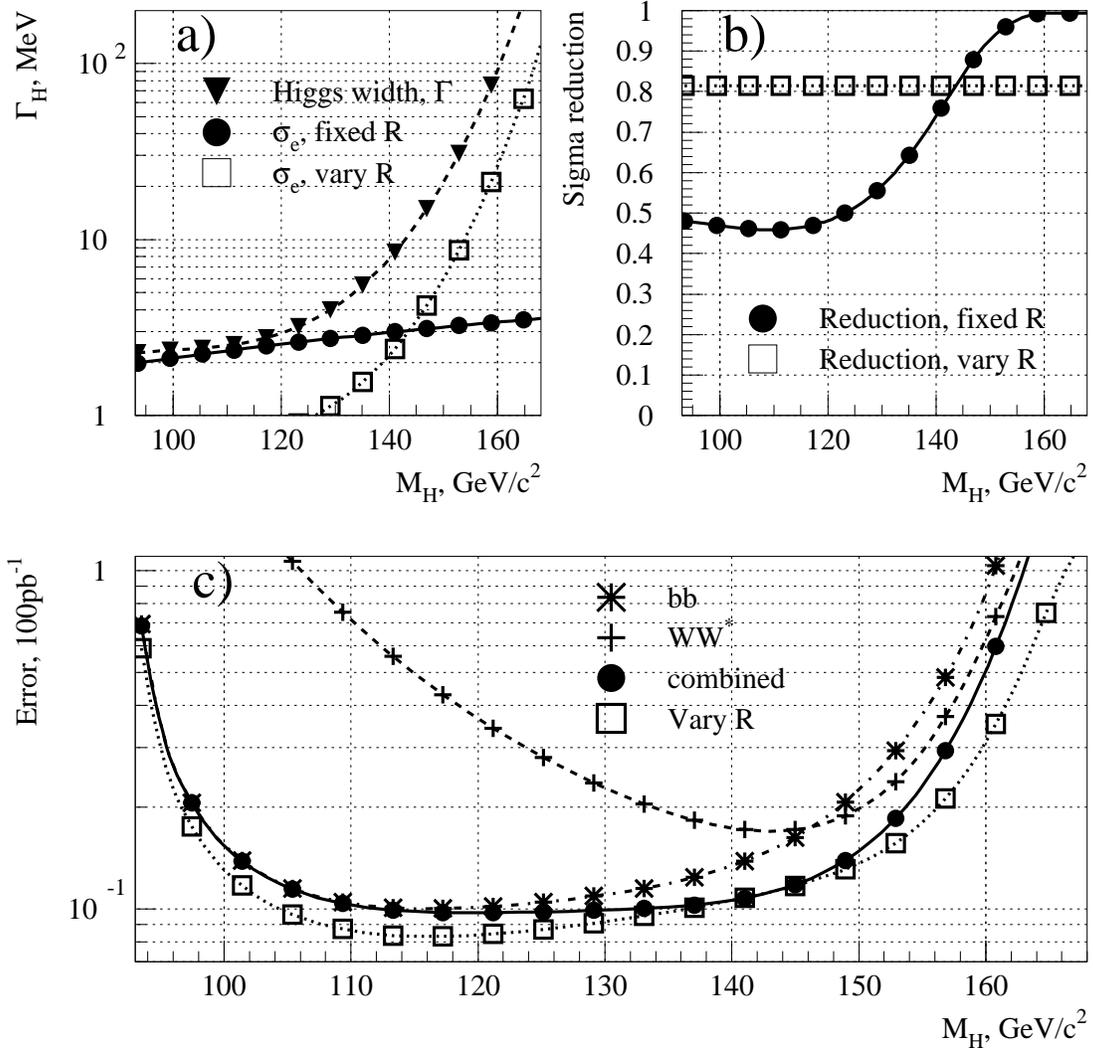,width=15cm,height=15cm}
\vspace*{-0.5cm}
\end{center}  
\caption[]{
Plot (a) shows the width of the SM Higgs boson as a function
of its mass (triangles), the centre-of-mass energy spread
for $R = 0.003\%$ (filled circles)
and the optimal varying energy spread (open squares). Plot (b)
shows the cross-section suppression factor due to
the width of the beams if $R = 0.003\%$ (filled circles), and for the 
optimal varying $R$ (open squares).
Plot (c) shows the fractional error with which the Higgs
cross section can be measured in the $b\overline{b}$ (stars) and 
$WW^*$ decay modes (crosses) using $100~pb^{-1}$ of data obtained 
with $R=0.003\%$. The solid circles shows the accuracy with which the 
peak cross section can be extracted if the SM branching ratios are 
assumed, and the open squares show the error obtained in the same 
running period by optimizing $R$.}
\label{fig:sigvmh}
\end{figure}

The width of the SM Higgs boson is shown as a function of its mass in 
Fig.~\ref{fig:sigvmh}(a), as a line with triangles. Also shown, 
with solid circles, 
is the spread in the centre-of-mass
energy for a collider with $R = 0.003\%$. 
The open squares correspond to the spread in the centre-of-mass energy
which is obtained if $R$ is varied so that the beam energy spread is 
always 40\% of the Higgs width. 
It is assumed here that any value of $R$ can be obtained, and that the
luminosity scales as $R^{2/3}$. This procedure approximately optimises 
the Higgs production rate, and hence the statistical error on the Higgs 
cross-section. Tighter beam energy spreads have lower luminosities, while 
increasing the spread reduces the Higgs cross-section.
Figure~\ref{fig:sigvmh}(b) shows how the reduction factor given in 
(\ref{peak-xs}) reduces the peak cross section in the two cases.

The decay mode $H\rightarrow b\overline{b}$ was investigated
in~\cite{cern99}, and those results are updated in
Fig.~\ref{fig:sigvmh}(c), taking account of the loss in peak cross
section. The suppression is less important as the mass, and hence the
width, rises. This means that the performance for $m_H = 140~\gevc$ is
almost the same as for $m_H = 115~\gevc$. We also display results for
the $WW^*$ decay mode, which is rather clean and has at least an 8\%
branching ratio in the SM. The accuracy of the width measurement
obtainable at a \mm\ collider in this channel is estimated by assuming
that the efficiency and background achieved by the DELPHI collaboration in
measuring $WW$ production at $161~\gevc$\cite{delphi-161} can be
duplicated. This includes the conservative assumption that the spin
information is not used to reduce the non-resonant $WW$ background. We
note that for $M_H = 115~\gevc$ a 6\% error on the $b\overline{b}$
cross-section and 32\% on the $WW^*$ are expected per $300~\pb^{-1}$,
or three years of running.

The decay mode $H\to\tau^+\tau^-$ is also an important channel, which
provides power to distinguish between different Higgs
models~\cite{Guasch}. The importance of this decay mode is discussed in
more detail in~\cite{Barger99} and in Sect.~\ref{sect:vercorr}, but we
recall here that a measurement of the branching ratio with a 16\%
statistical error could be made at a \mm\ collider using an integrated
luminosity of $100~\pb^{-1}$.

We conclude that if one varies $R$ at centre-of-mass energies above
$\sim$145~GeV, useful cross section measurements are possible up to about
160~GeV. Beyond this point, the Higgs resonance is simply too wide for a
peak cross-section measurement to be feasible.
However, we can confidently expect at least one Higgs boson in the 
mass range accessible to a \mm\ collider. 

The accuracies for the branching ratio measurements have to be compared 
with the corresponding numbers at an 
$e^+ e^-$ linear collider~\cite{tesla,orangebook,jlc}, where 
$\Delta BR/BR$ of about 2.5\%, 5\%, 4\% are achievable for the 
$b\bar b$, $\tau^+\tau^-$, $WW^*$ modes respectively (for $m_H=120~\gev$, 
$\sqrt{s}=350~\gev$, and $\int\!{\cal L}=500$~fb$^{-1}$).
We emphasize that the FMC accuracies quoted earlier for these modes are
very dependent on the luminosity obtainable, and that these LC numbers
include detailed detector simulations that are not yet available for the
FMC. In addition, by combining LC and FMC data the branching ratio of  
$H\to\mu^+ \mu^-$ can be measured to 4\% accuracy~\cite{Barger:2001mi}.

\begin{figure}[h!]
\begin{center}
\begin{picture}(148,58)
\put(0,0){\mbox{ \epsfig{file=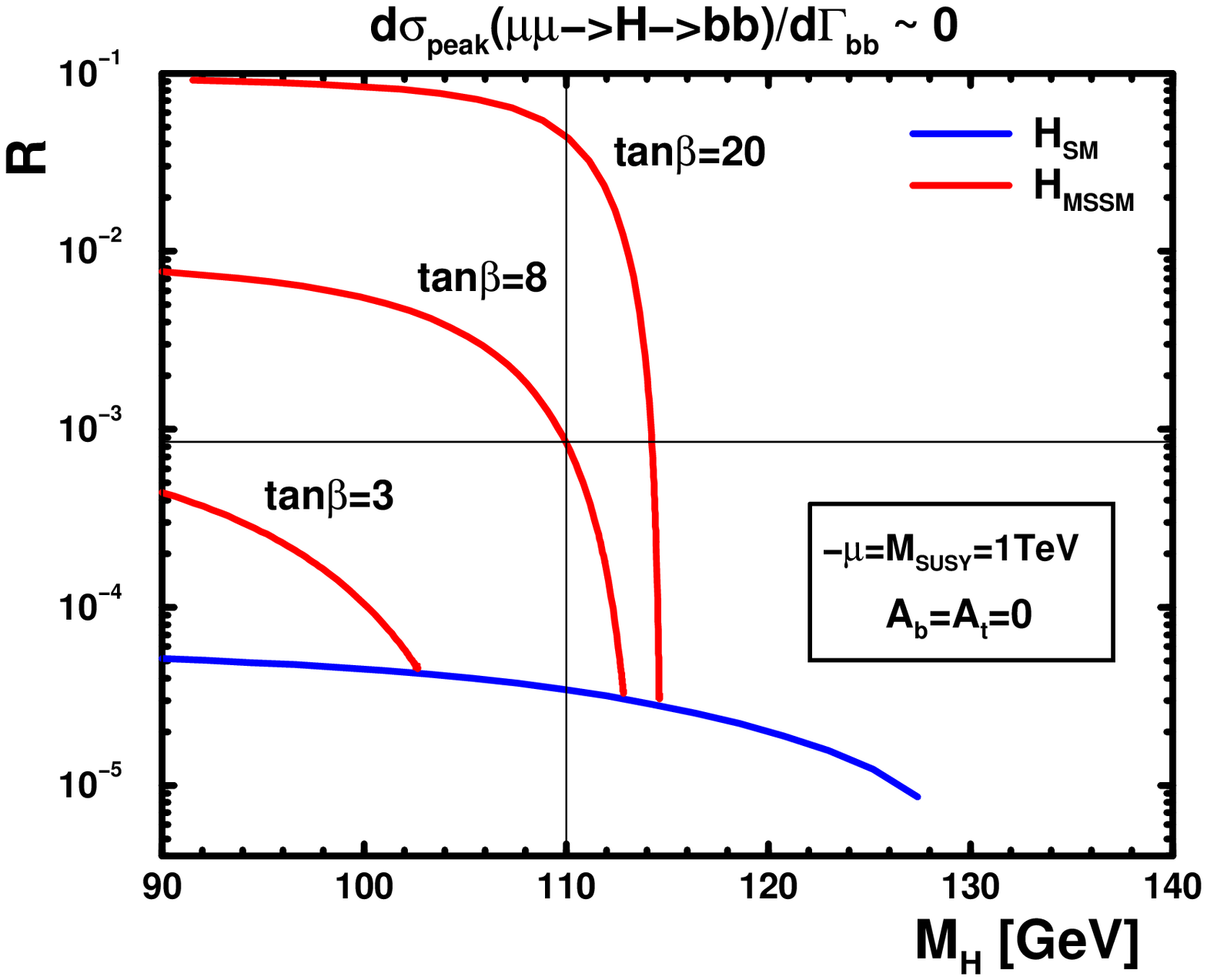, height=56mm} \hspace{8mm}
                 \epsfig{file=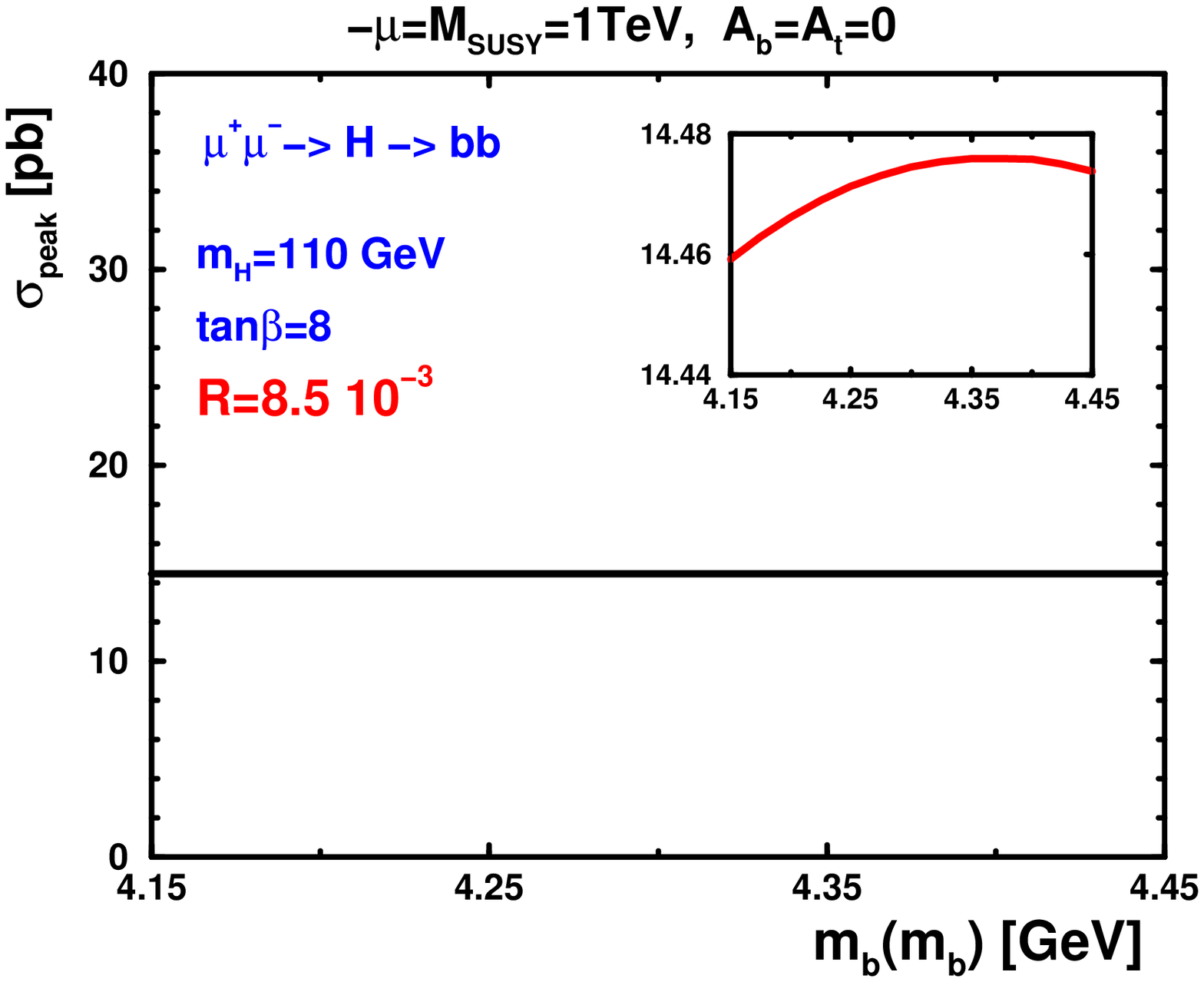, height=56mm} }}
\put(0,54){\mbox{\bf a)}}
\put(79,54){\mbox{\bf b)}}
\end{picture}
\end{center}
\vspace*{-5mm}
\caption{(a) Points in the $m_H$--$R$ plane where the $\mumu\to H\to\bb$
  cross section has a local maximum in the variable \mb. Curves are
  shown for the lightest neutral MSSM Higgs (assuming CP conservation) and
  for the SM Higgs. 
  In the former case, the relevant MSSM parameters are shown in the plot.
  The crossing vertical and horizontal lines pick up one such point
  for the MSSM Higgs. In (b), the \mb\ dependence of the peak
  cross section is shown for that point.  }
\label{fig:mbdep}
\end{figure}

It was observed in \cite{cern99} that, for certain values of $R$,
$\s_{\rm peak}(\mumu\to H\to\bb)$ becomes practically independent of
$m_b$.  More generally speaking, $R$ can be chosen such that the peak
cross section for a given final state $X$ is insensitive to $\GX$, i.e. 
$d\s_{\rm peak}/d\,\GX = 0$, or equivalently
\begin{equation}
\label{eq:stationary}
  \frac{\Gamma}{\GX}-1 = 
  \frac{2A}{\sqrt{\pi}}\, 
  \frac{e^{-A^2}}{1-{\rm Erf}(A)} - 2A^2 \,.
\end{equation}
In practice, this is only relevant if (i) $H\to X$ is the dominant
decay channel, $B(H\to X)>0.5$, and (ii) $\sigrts\lsim\GX$.  For both
$H_\mathrm{SM}$ and $H_\mathrm{MSSM}$ these conditions can be
simultaneously fulfilled only for $X=\bb$. 
Figure~\ref{fig:mbdep}(a) shows contours of 
$d\s_{\rm peak}/d\,\Gamma_{bb}=0$ 
in the $m_H$--$R$ plane for various values of $\tan\b$.  
Assuming a conservative error in the \MS\ bottom mass determination, 
namely $\omb(\omb)=4.30\pm 0.15$~GeV, would imply a 15\% indeterminacy
in the $\Gamma(H\to\bb)$ partial width.
If the accelerator parameters are tuned in such a way that
(\ref{eq:stationary}) is fulfilled, the impact of this uncertainty
on the theoretical prediction for the value of the $\mumu\to H\to\bb$
peak cross section is minimized. This is illustrated in
Fig.~\ref{fig:mbdep}\,b for a MSSM Higgs boson with a mass of 110~GeV,
for $\tan\b=8$, and $R=8.5\cdot 10^{-3}$.

In this way, the $\mumu\to H\to\bb$ 
peak cross section may be used as a precision measurement that may be 
interpreted, for example to constrain MSSM parameters as we discuss later, 
without any ambiguity related to \mb.

\subsection{\mbf Radiative Corrections to $\mumu\to h$}

Given the excellent experimental accuracy expected for a \mm\
collider, quantum corrections to the Higgs production and decay processes
have to be taken into account. For definiteness, we concentrate here on
the quantum corrections in the MSSM, assuming CP conservation, and leaving
the extension to include explicit CP violation induced by loop effects to
section~\ref{sect:CPviol} of this report.  We discuss how quantum
corrections affect the Higgs production cross sections and how they alter
the Higgs branching ratios.  Since these corrections depend on the
underlying supersymmetry parameters, precise measurements of the Higgs
boson masses, widths, branching ratios, {\it etc.} may be used to pin down
the parameter space of the MSSM. They may also be used for important
consistency checks of the model.

We focus on the three neutral Higgs bosons expected in the MSSM 
in the CP-conserving limit, namely the two scalars $h$ and $H$, 
and the pseudoscalar $A$.
Table~\ref{tab:hxcouplings} shows their tree-level couplings to
up- and down-type quarks/leptons and to the $W^\pm$ and $Z$ bosons,
relative to those of a SM Higgs boson. Here $\tan\b \equiv v_2/v_1$ and
$\alpha$ diagonalizes the CP-even Higgs sector,
$h=-\sin\a\,H_1^0+\cos\a\,H_2^0$, $H=\cos\a\,H_1^0+\sin\a\,H_2^0$.  In the
limit $m_A\to\infty$, $H$ and $A$ (and $H^\pm$) decouple from low-energy
physics, whilst $h$ becomes SM-like. At tree level, one has 
$m_h\to m_Z|\cos 2\b|$ and $\a\to\b-\pi/2$ in this case.

\begin{table}[h!]
\begin{center}
\begin{tabular}{|c|c|c|c|} 
  \hline
        & $t$ & $b,\,\tau\,(\mu)$ & $W,Z$ \\
  \hline
  $h^0$ & $\cos\alpha/\sin\beta$ & $-\sin\alpha/\cos\beta$ &  $\sin(\beta-\alpha)$ \\
  \hline
  $H^0$ & $\sin\alpha/\sin\beta$ & $\cos\alpha/\cos\beta$ & $\cos(\beta-\alpha)$\\
  \hline
  $A^0$ & $-i\gamma_5\cot\beta$ & $-i\gamma_5\tan\beta$ &  0 \\
  \hline
\end{tabular}
\end{center}
\caption{Tree-level couplings of the MSSM Higgs bosons $h$, $H$, and $A$ 
         relative to those of a SM Higgs boson.}
\label{tab:hxcouplings}
\end{table}

The CP-even Higgs mass matrix ${\cal M}^2$ is, however, subject to large 
radiative corrections, the leading effects being 
$\propto h_t^4$~\cite{mhiggscorr}, with $h_t^{}$ the top Yukawa coupling.  
A one-loop renormalization-group- (RG-)improved effective potential 
calculation gives \cite{CMW} 
\begin{eqnarray} 
   {\cal M}_{12}^2 &\sim& 
   - \left[ m_A^2 + m_Z^2 - {h_t^4 v^2 \over 8\pi^2\MSUSY^2}\: 
            \mu^2\left(3-{A_t^2\over \MSUSY^2}\right) \right] \sin\b\cos\b
   \nonumber \\[2mm]
   & & +\; \xi \left[ 
     {h_t^4 v^2 \over 16\pi^2}\,\sin^2\!\b\:{\mu A_t\over \MSUSY^2} 
     \left({A_t^2\over \MSUSY^2}-6\right) + 3\, {h_t^2m_Z^2\over 32\pi^2}\,
     {\mu A_t\over \MSUSY^2} \right]
  \label{eq:M12}
\end{eqnarray}
where $\xi$ accounts for the leading-logarithmic two-loop effects,
\MSUSY\ is an overall supersymmetry scale, and $A_t$ is the trilinear
soft SUSY-breaking parameter in the stop mass matrix. Diagonalizing
the loop-corrected ${\cal M}^2$ gives the higher-order values of
$m_h$, $m_H$, and the effective mixing angle $\aeff$, with
\begin{equation}
  \sin 2\aeff = \frac{ 2{\cal M}_{12}^2 }{
       \sqrt{({\rm Tr}{\cal M}^2)^2-4\,{\rm det}\,{\cal M}^{\,2} }} \,.
\end{equation}
As has been shown analytically in \cite{hff}, if supersymmetric vertex
corrections are neglected, the improved tree-level couplings are
obtained by replacing $\a\to\aeff$ in Table~\ref{tab:hxcouplings}.
Note, however, that the vertex corrections may well be important, 
as is discussed for the $hb\bar b$ and $h\tau^+\tau^-$ vertices in
Sect.~\ref{sect:vercorr}.

While in a large fraction of the MSSM parameter space 
the couplings of $h$ to $\bb$, $\mumu$, $\tautau$ are 
enhanced compared to those of a SM Higgs boson, it is, 
according to (\ref{eq:M12}), also possible to have 
${\cal M}_{12}^2\to 0$, which corresponds to $\sin\aeff\to 0$ 
or $\cos\aeff\to 0$. 
Thus $g_{hbb}$, $g_{h\tau\tau}$ or $g_{h\mu\mu}$ can be strongly
reduced in the MSSM~\cite{CMW,redcoup}. 
This could lead to a highly suppressed production
cross section for the lightest CP-even Higgs boson, 
even if it is in the kinematically accessible region.

\begin{figure}[htb!]
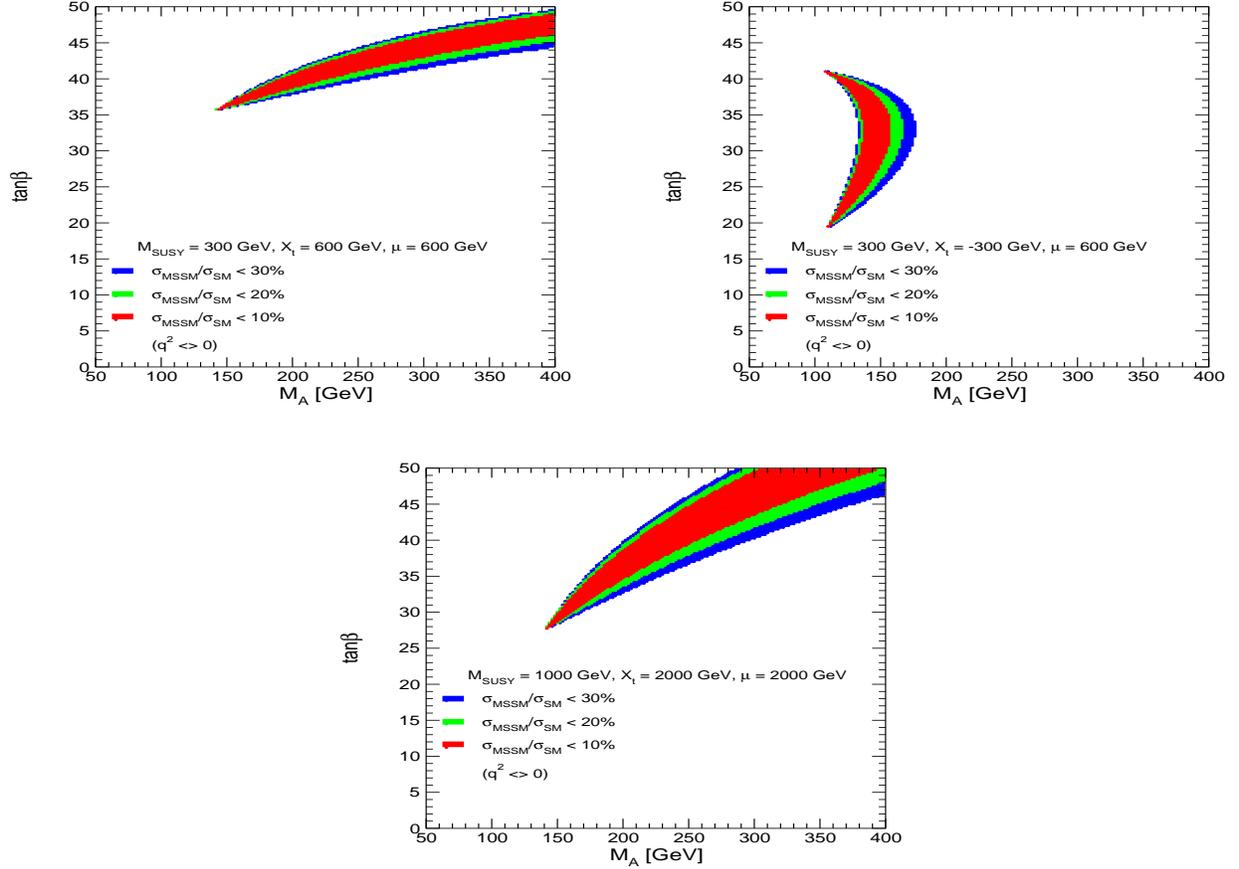

\begin{center}
\epsfig{figure=muC36.cl.eps,width=7.5cm,height=5.4cm}
\hspace{1cm}
\epsfig{figure=muC33.cl.eps,width=7.5cm,height=5.4cm}
\vspace{0.7cm}

\epsfig{figure=muC56.cl.eps,width=7.5cm,height=5.4cm}
\end{center}
\caption[]{Regions in the $m_A$--$\,\tb$ plane where 
  $\sigma_{\rm MSSM}(\mu^+\mu^- \to h)/\sigma_{\rm SM}(\mu^+\mu^- \to
  h) < 30\%$ for a common supersymmetric mass scale 
  $\MSUSY = 300$~GeV (top) and $\MSUSY = 1000$~GeV (bottom).}
\label{fig:sigmasmall}
\end{figure}

We display in Fig.~\ref{fig:sigmasmall} the regions of parameter space
where $\sigma_{\rm MSSM}(\mu^+\mu^- \to h)/\sigma_{\rm SM}(\mu^+\mu^- \to
h) < 30\%$ (for details, see~\cite{mumuhsuppression}). Results obtained
with the {\tt FeynHiggs} code~\cite{feynhiggs} are displayed in the
$m_A$--$\,\tb$ plane for $\MSUSY = 300,1000 {\rm ~GeV}$. The off-diagonal
entry in the $\tilde t$ mass matrix is taken as $X_t \equiv A_t - \mu/\tb
= \pm\MSUSY,\,2\,\MSUSY$. Moreover, we choose $\mu = 2\,\MSUSY$ and $A_b =
A_t$.  The other MSSM parameters are $M_2=400 {\rm ~GeV}$ and $m_{\tilde
g}=\MSUSY$, and we take $m_t = 175 {\rm ~GeV}$ 
and $\omb=\omb(m_t)=2.97 {\rm ~GeV}$.

The top plots in Fig.~\ref{fig:sigmasmall} show the case with a relatively
small supersymmetry mass scale, $\MSUSY = 300 {\rm ~GeV}$, for the two
combinations $X_t = \mu = 600$~GeV and $X_t = -300$~GeV, $\mu = 600$~GeV.
In general, the regions with $\sigma(\mu^+\mu^- \to h) < 30\%$ are
obtained for large $\tb \gsim 15$ and relatively small $m_A$: $100 {\rm
~GeV} \lsim m_A \lsim 300 {\rm ~GeV}$.  This follows from (\ref{eq:M12})
in the limit ${\cal M}_{12}^2 \to 0$, since $m_A$ has to be relatively
small in order for the higher--order corrections to be of a size similar
to $(m_A^2 + m_Z^2)\sin\beta \cos\beta$. Due to the functional dependence
of ${\cal M}_{12}^2$ on $\mu$ and $X_t$ $\approx A_t$ for $\tb
\gsim 10$ as seen in (\ref{eq:M12}), the regions hardly change their shape
if both $X_t$ and $\mu$ change their sign. However, their shape and
location change drastically if only one sign flips.  In the bottom plot in
Fig.~\ref{fig:sigmasmall}, we show the corresponding regions for a large
supersymmetry mass scale, $\MSUSY = 1000$~GeV, and $X_t = \mu = 2000$~GeV.
As in the low \MSUSY\ case, we find a non-negligible part of parameter
space with a highly suppressed $\sigma(\mu^+\mu^- \to h)$~\footnote{The
regions in Fig.~\ref{fig:sigmasmall} with suppressed production cross
section are not significantly affected by the exclusion bounds obtained
from the Higgs search at LEP~\cite{LEPtb,lepHiggs}.}.
A more extensive discussion can be found in
Ref.~\cite{mumuhsuppression}.

\begin{figure}[htb!]
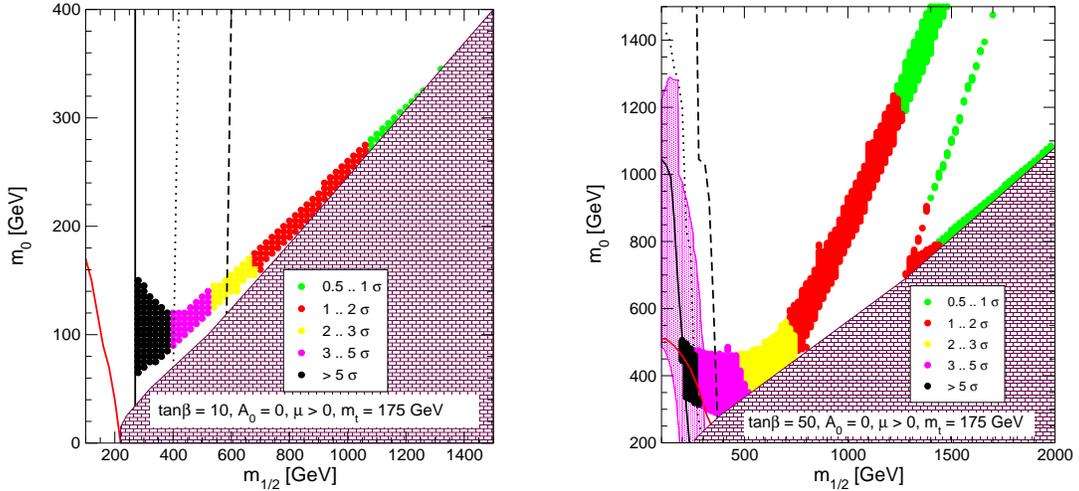

\begin{center}
\epsfig{figure=EHOW31e.03.cl.eps,width=6.5cm,height=6.5cm}
\hspace{1cm}
\epsfig{figure=EHOW31e.09.cl.eps,width=6.5cm,height=6.5cm}
\end{center}
\caption[]{ 
$[\sigma(\mm \to h) \times {\rm B}(h \to b \bar b)]_{\rm CMSSM}$ 
compared to the SM value
in the $m_{1/2}-m_0$ plane for $\mu > 0$ and 
$\tan\beta = 10, A_0 = 0$ (left plot) and 
$\tan\beta = 50, A_0 = -2 m_{1/2}$ (right plot)~\cite{ehow2},  
assuming an experimental accuracy of 3\%.  
The bricked region is forbidden because the LSP is the lightest 
$\tilde\tau$. The regions above and to the right of the (red) diagonal 
solid lines yield values of $g_\mu - 2$ within $2 \sigma$ of the 
present central value. 
The light shaded (pink) region is excluded by  
${\rm B}(b \to s \gamma)$ measurements. The solid leftmost  
(dotted middle, dashed rightmost) near-vertical line corresponds 
to $\mh = 113$ (115, 117)~GeV~\cite{feynhiggs}.}
\label{fig:ehow}
\end{figure}

The parameter regions where $\sigma(\mu^+\mu^- \to h)$ goes to zero are
clearly somewhat unusual, and this possbility is not realized in the
constrained MSSM (CMSSM), in which the soft supersymmetry-breaking scalar
masses $m_0$ and gaugino masses $m_{1/2}$ are required to be universal at
an input GUT scale~\cite{ehow2}. This is exemplified in
Fig.~\ref{fig:ehow}, where we show 
$[\sigma(\mm \to h) \times {\rm B}(h \to b \bar b)]_{\rm CMSSM}$ 
in terms of standard deviations from the SM value
in the $m_{1/2}-m_0$ plane for $\mu > 0$ and two combinations
of $\tan\beta$ and $A_0$, together with the constraints from 
${\rm B}(b \to s \gamma)$~\cite{bsgexp,bsgtheo}, 
$g_\mu - 2$~\cite{gminus2} and
the requirement that the LSP is uncolored and uncharged~\cite{EHNOS}.
The mass of the lightest MSSM Higgs boson is also 
indicated~\cite{feynhiggs}.  
We have assumed an accuracy of 3\% \cite{cern99} in the 
determination of $\sigma(\mm \to h) \times {\rm B}(h \to b \bar b)$.
No suppression of Higgs production can be
observed for the CMSSM parameter space. On the contrary, the
production and decay is always enhanced compared to the
corresponding SM value, over the entire CMSSM parameter space~\cite{ehow2}. 
However, the existence in principle of the regions in the
unconstrained MSSM with suppressed Higgs production cross section
does point up the interest of measuring this observable.

\begin{figure}[htb!]
\begin{center}
\centerline{\epsfig{file=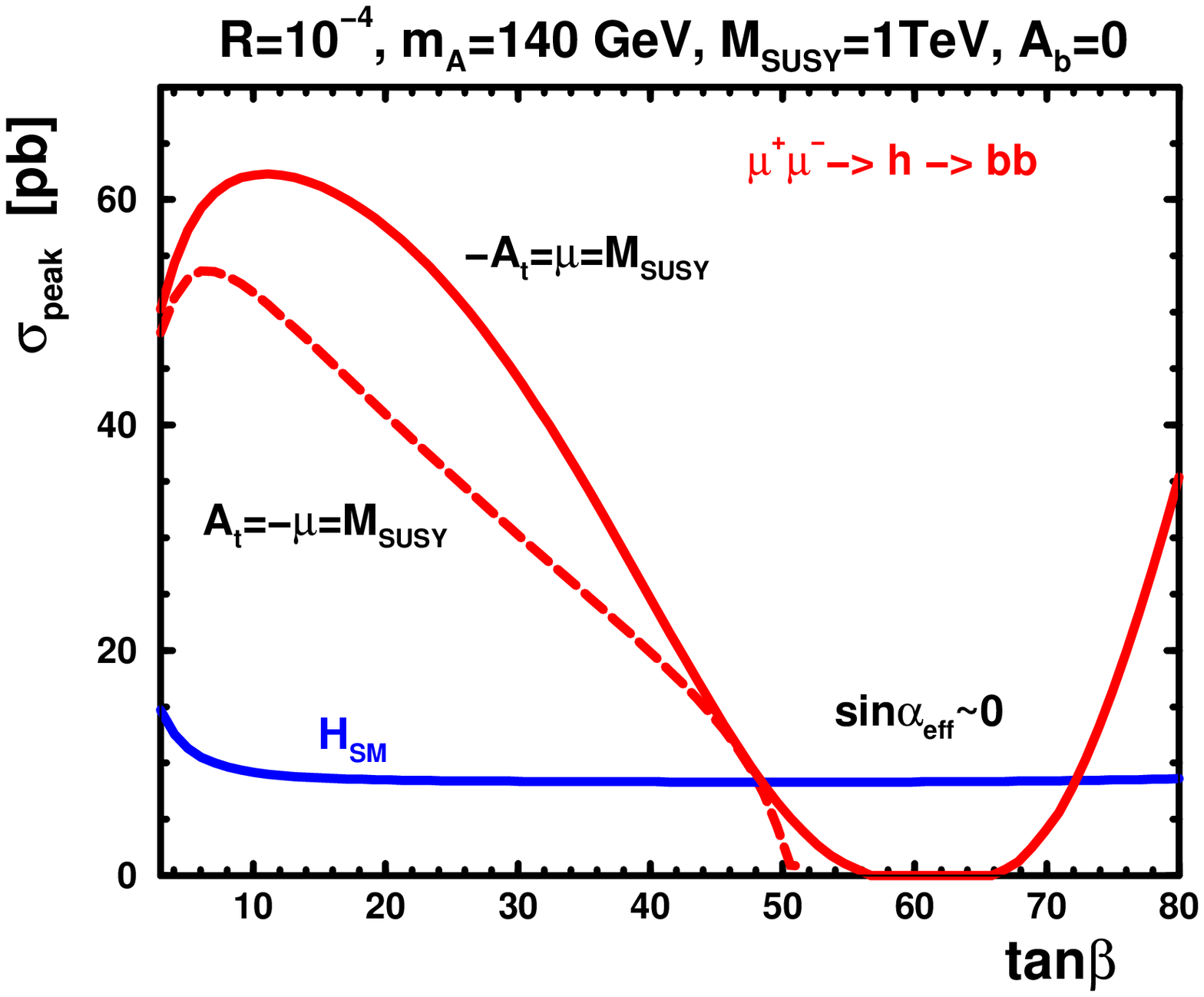,height=56mm} \hspace{8mm}
            \epsfig{file=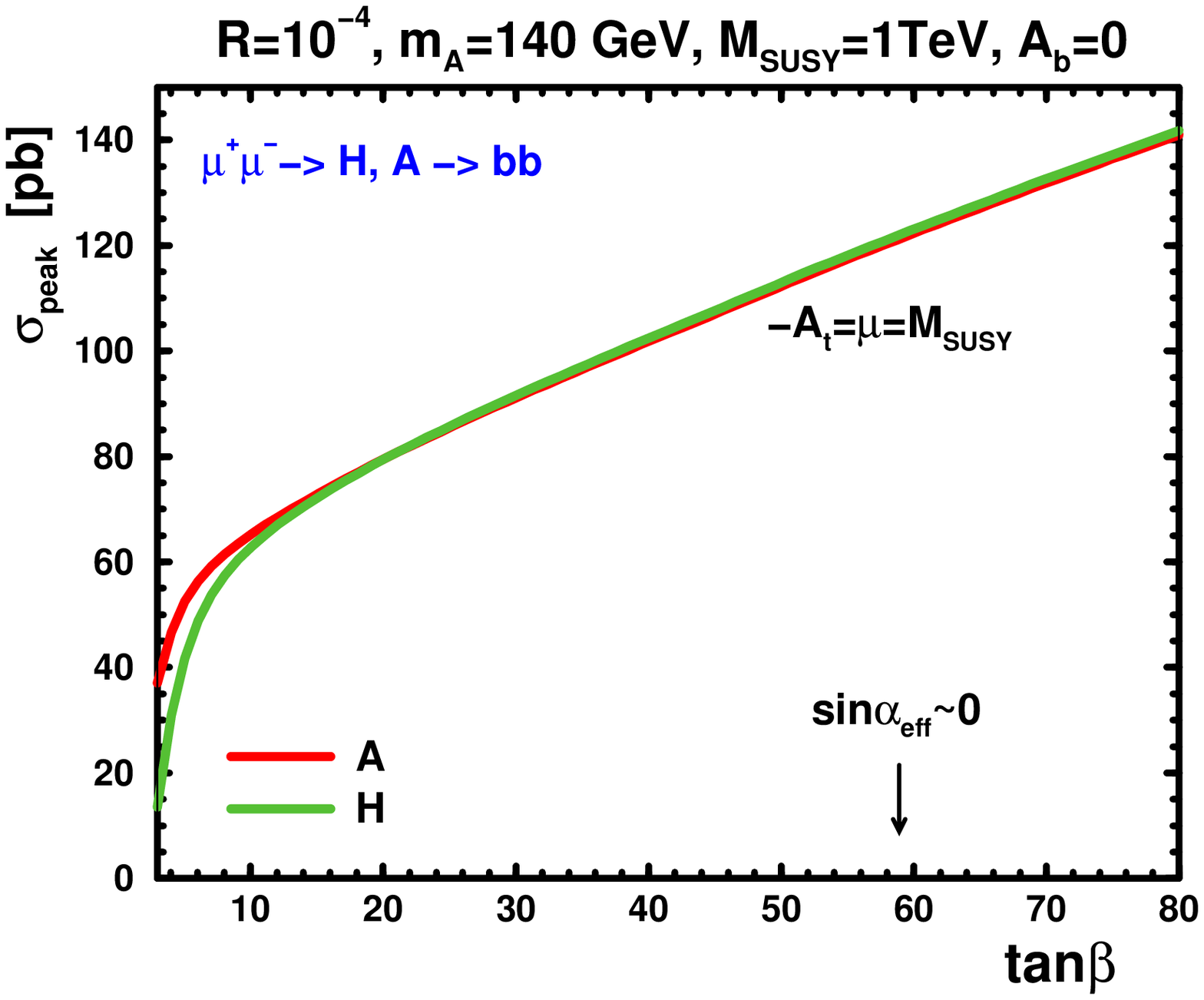,height=56mm} }
\end{center}
\vspace*{-1cm}
\caption{The $\tan\b$ dependence of the $\mumu\ \to (h,H,A)\to\bb$ peak
cross sections. The $H_{SM}$ curve has a small dependence on  $\tan\b$ on
the left, reflecting the fact that we set
the SM Higgs mass equal to $M_h$.
}
\label{fig:sina}
\end{figure}

We note also that, even when $\sigma(\mu^+\mu^-\to h)$ does vanish, 
the production cross sections of $H$ and $A$ are unsuppressed and 
even enhanced by $\tb$. Moreover, $H$ and $A$ would  
mainly decay into $\bb$ or $\tautau$.  We find that in this case the
usual mass hierarchy between $H$ and $A$ is inverted: The CP--odd
Higgs-boson mass $m_A$ turns out to be larger than $m_H$ by up to $25 {\rm
~GeV}$.  For a considerable fraction of the parameter regions with
suppressed $\sigma(\mu^+\mu^- \to h)$, the mass splitting $m_A - m_H$ is
larger than the sum of the total widths of $A$ and $H$ (this holds in
particular for not too large values of $\tb$).  Thus, in these regions $A$
and $H$ should not only be produced with sufficiently high rates but
should also be resolvable as separate resonances. Even in the regions
where $\sigma(\mu^+\mu^- \to h)$ is suppressed, the \mm\ collider would
therefore possess a promising potential for probing the neutral MSSM Higgs
sector via resonant production of $H$ and $A$.  This is illustrated in
Fig.~\ref{fig:sina}, where the peak cross sections of $\mumu\to \{h,\,H\,
{\rm or}\, A\}\to\bb$ are shown as a function of $\tan\b$ for
$m_A=140$~GeV and $M_{\rm SUSY}=\mu=-A_t=1$~TeV.  
Note that, for a large range of values of $\tan\b$, the 
$\mumu\to h\to\bb$ cross section is much larger than
that of a SM Higgs boson of the same mass.  For $\tan\b\sim 60$, however,
$\sin\aeff\sim 0$ and $\sigma(\mu^+\mu^- \to h)$ vanishes.  On the other
hand, $\s_{\rm peak}(\mumu\to H,A\to\bb)$ increases almost linearly for
$\tan\b\gsim 10$.

\subsection{\mbf Corrections to the $h\bb$ and $h\tautau$ Vertices}
\label{sect:vercorr}

An interesting topic that could be investigated at a \mm\ collider is
the effect of supersymmetric threshold corrections on the Yukawa
interactions. For large \tb\ and $\mu$ one expects
relative deviations of order unity of the couplings from their
tree-level values, due to gluino (SUSY-QCD) and, to a lesser extent,
higgsino (SUSY-EW) radiative effects.\footnote{Large-\tb\ scenarios, 
  like those derived from some supersymmetric SO(10) models 
  \protect{\cite{gut,Dmb}}, have become more appealing since LEP
  searches for a light neutral Higgs boson started excluding the
  low-\tb\ region of the MSSM parameter space
  \protect{\cite{lepHiggs}}.} 
Such deviations would be strongly correlated, as discussed below,
allowing for non-trivial self-consistency checks of the model.
Moreover, these corrections should of course be taken into account
when determining the underlying SUSY parameters from the Higgs bosons 
production cross sections and branching ratios.

\begin{figure}[htb]
\begin{center}
\centerline{\epsfig{file=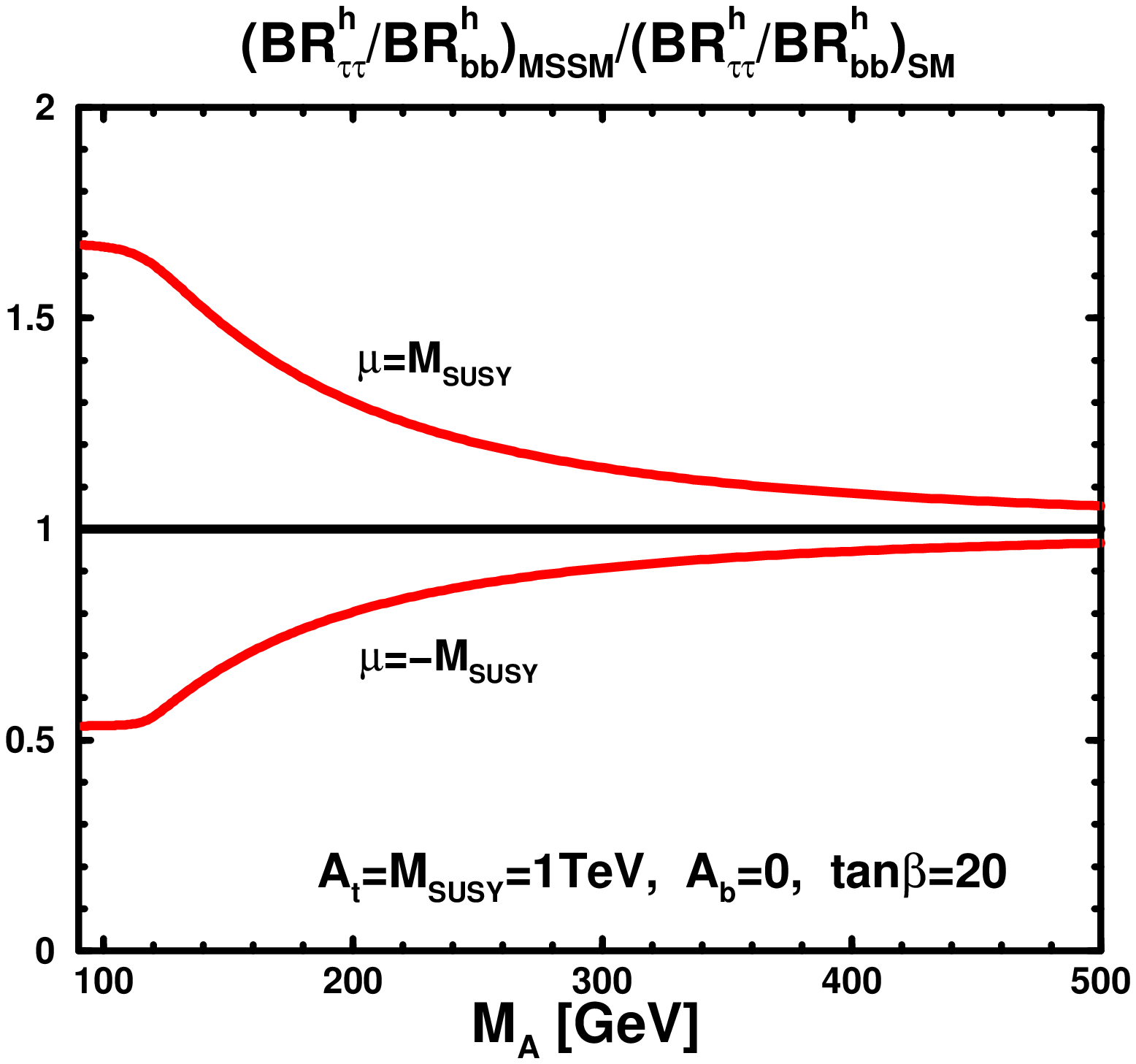,height=56mm}}
\end{center}
\vspace*{-1cm}
\caption{Ratio of the $h\to\tautau$ and $h\to\bb$ branching ratios in the 
MSSM relative to those in the SM.}
\label{fig:HbtauBRs}
\end{figure}

It can be proven that, in mass-independent renormalization schemes
like \MS, all-order SUSY corrections of the form
$\aS[n]\tan^n\!\b$ can be resummed in the following definition for
\hb~\cite{cgnw}:
\begin{equation}
  \label{eq:hb}
  \hb\,v_1=\frac{m_b}{1+\Delta\hb^1/\hb+\Delta\hb^2/\hb\,\tb}
                      \sim\frac{\mb}{1+\Dmb}\,,
\end{equation}
with the quantity \Dmb\ being dominated by SUSY-QCD virtual
effects:\footnote{A similar formula can be written for the $\tau$
  Yukawa coupling, although the SUSY-EW contributions in $\Delta\mtau$ 
  are (generally) smaller.}
\begin{equation}
\label{eq:Dmb}
  \Dmb\sim\Dmb[\mathrm{SQCD}]=
  \frac{2\aS}{3\pi}\,
       \mu\mg\,\tb\,I(m_{\tilde{b}_1},m_{\tilde{b}_2},m_{\tilde{g}})\,,
\end{equation}
where $I$ is the limit of the Passarino-Veltman function $C_0$ for 
vanishing
external momenta. It can be shown that \Dmb\ does not vanish in the
limit $\MSUSY\to\infty$. In fact, it approaches
$\aS/(3\pi)\,\tb$ asymptotically. This should not be understood as a 
non-decoupling
effect, as its physical consequences in low-energy observables do
vanish in the limit $\mA\to\infty$, when the SM is recovered as an
effective theory, as discussed below.
There is, however, an important phenomenological consequence of the 
non-vanishing
behaviour of~(\ref{eq:Dmb}): even for a very heavy supersymmetric
spectrum (but not too large $\mA$), one expects large deviations from
the normal ratio 
\begin{equation}
  g_{hbb}/g_{h\tau\tau}=\mb/\mtau\,,
\end{equation}
which holds not only in the SM, but also in two-Higgs-doublet models 
(2HDMs) of types I and II \cite{twoHDM}. 
This translates for instance, into the corresponding ratio of branching
fractions, as can be seen in Fig.~\ref{fig:HbtauBRs}. 
Since this ratio may 
be measurable with a precision better than 16\%, as mentioned previously, 
multi-standard-deviation discrimination between the MSSM and the SM is 
possible for small $m_A$.

\begin{figure}[htb]
  \centerline{\epsfig{file=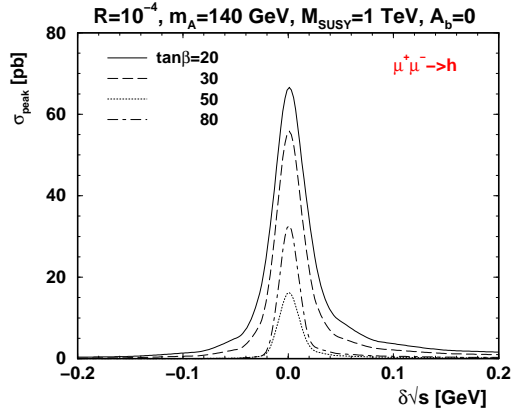,height=56mm}}
\vspace*{-4mm}
\caption{Resonant $\mumu\to h$ cross sections, for
    $\tb=20,\,30,\,50,\,80$, as a function of the distance to the peak of 
    the resonance, $\delta\sqrt{s}$. The remaining MSSM parameters are 
chosen as in Fig.~\protect{\ref{fig:sina}}.}
\label{fig:mmh9}
\end{figure}

Using (\ref{eq:hb}), and after adding the process-dependent
SUSY vertex corrections,
the renormalized amplitudes for $h\to b\bar{b}$, $H\to b\bar{b}$ and 
$A\to b\gamma_5\bar{b}$ read \cite{CMW}: 
\begin{eqnarray}
\label{eq:renamp}
{\cal A}(\bar{b}b\,h)
  &\sim & \frac{\mb}{v}\,\frac{\sin\aeff}{\cbt}   \frac{1}{1+\Dmb}
      \left(1-\frac{\Dmb}{\tan\aeff\tb}\right)\,, \\[1.5ex]
{\cal A}(\bar{b}b\,H)
  &\sim &-\frac{\mb}{v}\,\frac{\cos\aeff}{\cbt}   \frac{1}{1+\Dmb}
    \left(1+\Dmb\,\frac{\tan\aeff}{\tb}\right)\,, \\[1.5ex]
{\cal A}(i\,\bar{b}\gamma_5 b A)
  &\sim & \frac{\mb}{v}\,{\tb} \frac{1}{1+\Dmb}\,.
\label{eq:renamp-e}
\end{eqnarray}
As can be checked in~(\ref{eq:renamp}) one recovers the SM
coupling for $h b\bar{b}$ (and $G^0 b\bar{b}$) as $\mA/M_W\to\infty$,
because $\tan\aeff\to -1/\tb$. 
We see in (\ref{eq:renamp}) how a radiative zero of the \mmhbb\ 
cross section could occur if 
\begin{equation}
  \tan\aeff=\frac{\Dmb}{\tb}\ \ \Longrightarrow\ \ g_{hbb}\sim 0 \,.
\end{equation}
At the same time, $g_{h\mu\mu}$ and $g_{h\tau\tau}$ are not zero but 
\begin{equation}
  g_{h\tau\tau(\mu\mu)} = {m_{\tau(\mu)}\over v}\:
        {\Delta m_b - \Delta m_{\tau(\mu)} \over 1+\Delta m_{\tau(\mu)}} \,.
\end{equation}
For $\Delta m_b\sim {\cal O}(1)$, this is of the same order as the 
corresponding coupling in the SM. 
Hence $\s(\mumu\to h)$ is not suppressed, but $h$ does not decay into
$\bb$. Instead, it mainly decays into $\tautau$ or $WW^*$. 
The total decay width, however, becomes very small, of the order of a few
MeV, leading to an extremely narrow resonance even for large $\tan\b$. 
An example is shown in Fig.~\ref{fig:mmh9}, where we plot 
$\s_{\rm peak}(\mumu\to h)$ as a function of $\delta\rts\equiv\rts-m_h$ 
for various $\tb$ values. 
Notice that, for small $\Gamma_h$ values, the actual width of the
resonances is approximately given by $R\sqrt{s} \sim \sigrts$ (see
(\ref{eq:sigapprox2})).

Last but not least,  
it should be mentioned that the treatment of the 
vertex corrections in terms of $\Delta m_{b(\tau)}$ is a very good 
approximation for large $\tan\b$ and $M_{\rm SUSY}\gg m_A$. 
For $M_{\rm SUSY} \sim m_A$, however, this should be complemented with 
the remaining neglected terms in a full one-loop
computation~\cite{hff,maria}. 
Here the problem of a consistent treatment of the stop and sbottom 
system arises. For a
prescription which includes a consistent renormalization of the 
$\ti t$ and $\ti b$ sectors and which is, moreover, valid for all 
allowed $\tb$ values, see~\cite{ehkmy}.

\section{ CP Violation at \mbf $\mu^+\mu^-$ Colliders} 
\label{sect:CPviol}

\setlength{\unitlength}{1pt}

\subsection{Introduction}

In this Section, we discuss ways in which models of CP violation may be
probed at both the FMC and the SMC. We recall that, although the effective
Higgs potential of the Standard Model (SM) conserves CP, at least up to
the two-loop level, this symmetry is generically violated in models which
extend the Higgs sector even minimally, such as the two-Higgs-doublet
model~\cite{TDLee}. Very interestingly, even though CP symmetry can be
imposed on the complete Lagrangian of a three-Higgs-doublet model, it can
still be broken spontaneously by the Higgs ground state \cite{SW}.  In
supersymmetric theories, CP violation may be generated either
spontaneously~\cite{GP,JCR} or explicitly~\cite{APLB,CEPW} via loop
effects. In particular, the MSSM with explicit radiative CP breaking in
the Higgs sector constitutes a prototype scenario for studying CP
violation at a $\mu^+\mu^-$ machine. As we argue below, the option of muon
polarization at a \mm\ collider would be a particularly valuable tool for
determinating the CP properties of Higgs bosons.

In this connection, we first review briefly the basic mechanism for
obtaining resonantly-enhanced CP asymmetries in Higgs-mediated processes
at the FMC. Then we discuss the CP-violating effects due to vertex
corrections in the MSSM. Next we define optimal CP-violating observables
based on muon polarization and give estimates of the expected CP violation
in the MSSM and other extended scenarios.  We also summarize the
requirements for reducing the CP-conserving background. Finally, we give
an example of the possible impact of CP violation on the mass difference
between the second and third neutral Higgs bosons in the MSSM with CP
violation~\cite{CEPW}.

\subsection{Resonant CP Violation due to Higgs-Mass Mixing}

\begin{figure}[t]
\begin{center}
\begin{picture}(400,100)(0,0)
\SetWidth{0.7}

\ArrowLine(0,100)(50,50)\ArrowLine(0,0)(50,50)\GCirc(50,50){10}{0.3}
\DashLine(50,50)(150,50){5}\ArrowLine(150,50)(200,100)\ArrowLine(150,50)(200,0)
\GCirc(50,50){10}{0.6}\GCirc(100,50){15}{0.8}\GCirc(150,50){10}{0.6}
\Text(-5,100)[r]{\color{Blue}$a$}\Text(-5,0)[r]{\color{Blue}$b$}
\Text(205,100)[l]{1}\Text(205,0)[l]{2}
\Text(70,65)[b]{${\color{OliveGreen}H},{\color{Red}A}$}
\Text(130,65)[b]{${\color{OliveGreen}H},{\color{Red}A}$}

\ArrowLine(300,100)(350,100)\ArrowLine(350,100)(400,100)
\ArrowLine(300,0)(350,0)\ArrowLine(350,0)(400,0)
\GOval(350,50)(50,15)(0){0.7}
\Text(295,100)[r]{\color{Blue}$a$}\Text(295,0)[r]{\color{Blue}$b$}
\Text(405,100)[l]{1}\Text(405,0)[l]{2}
\end{picture}
\end{center}
\caption{Resonant and non-resonant Higgs contributions to a generic
process $ab \to 1 2$.}\label{f1} 
\end{figure}
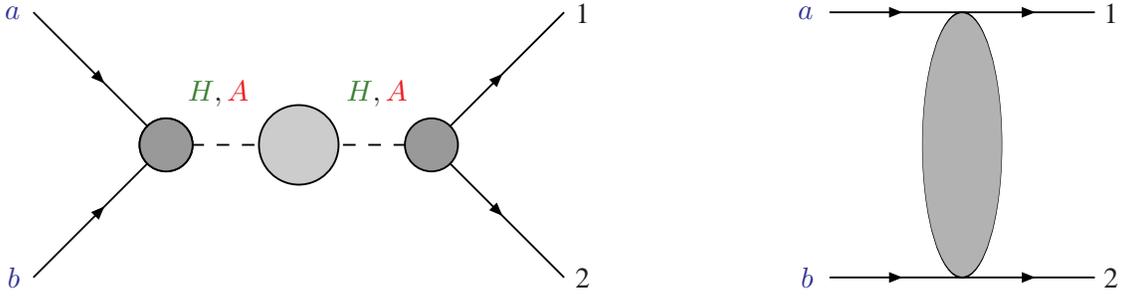

The general formalism for mass mixing in extended Higgs sectors with
explicit CP violation induced by loop effects is well
developed~\cite{ANPB,APLB}. It is instructive for our purposes here to
recall the conditions necessary for CP asymmetries to exhibit resonant
enhancement. For this purpose, we consider a generic process $ab \to
12$, as shown in Fig.~\ref{f1}.  Such a reaction may proceed via both
CP-even and CP-odd Higgs bosons, $H$ and $A$, as intermediate states.
The corresponding transition amplitude may be written as
\begin{equation}
  \label{Tamp}
{\cal T}\ =\ {\cal T}^{res}\, +\, {\cal T}^{\rm box}\ =\ V^P_i \left(
\frac{1}{s\, -\, {\cal H}(s)}\right)_{ij} V^D_j\ +\ {\cal T}^{\rm box}\, ,
\end{equation}
where 
\begin{equation}
  \label{InvDHA}
s\, -\, {\cal H}(s)\ =\ \hat{\Delta}^{-1} (s)\ =\ s\mbox{\bf 1}\, -\,
\left[
\begin{array}{cc}
M^2_A-\widehat{\Pi}^{AA}(s) & -\widehat{\Pi}^{AH}(s)\\
-\widehat{\Pi}^{HA}(s) & M^2_H - \widehat{\Pi}^{HH}(s)
\end{array} \right]\, 
\end{equation}
is the inverse-propagator matrix, which describes the dynamics of the
$H-A$ mass mixing.\footnote{For a full discussion, including
three--state $h-H-A$ mixing, see \cite{cepw3}.}  The propagator matrix
$\hat{\Delta}(s)$ actually arises from summing a geometric series of
$HH$, $AA$, $HA$ and $AH$ self-energies.  In~(\ref{InvDHA}), the hat
symbol $\:\widehat{~}\:$ denotes the fact that the resummed self-energies
should be renormalized within a gauge-invariant resummation approach
that respects unitarity~\cite{ANPB}, e.g., the one implemented by the
Pinch Technique (PT)~\cite{PP}.

Assuming that the non-resonant term ${\cal T}^{\rm box}$ in
(\ref{Tamp}) is negligible, we may identify two sources of CP
violation: (i) the non-vanishing scalar-pseudoscalar mixing term
$\widehat{\Pi}^{AH}(s)$ in the effective Higgs-boson mass matrix, and
(ii) CP violation in the production and decay vertices $V^P_i$ and
$V^D_i$. By analogy with the neutral kaon system, we term these
$\varepsilon$- and $\varepsilon'$-type effects, respectively. We first
discuss the $\varepsilon$-type effects, which will be resonantly
enhanced if~\cite{ANPB}
\begin{equation}
  |M^2_H - M^2_A - \widehat{\Pi}^{HH}(s) + \widehat{\Pi}^{AA}(s)|\ 
  \stackrel{<}{{}_\sim}\ 2|\widehat{\Pi}^{HA}(s)|\, .
  \label{CPres}
\end{equation}
The condition~(\ref{CPres}) is naturally  fulfilled in models  where the
CP invariance  of  the Higgs potential  is minimally  lifted by radiative
effects, such as the MSSM~\cite{APLB,CEPW} and heavy Majorana-neutrino
models inspired by E$_6$ theories~\cite{ANPB}.

\begin{figure}[ht]
\begin{center}
\begin{picture}(400,180)(0,0)
\SetWidth{0.7}

\DashArrowLine(0,100)(30,100){5}\Text(10,110)[br]{{\color{Red}$a$}}
\DashArrowLine(100,100)(130,100){5}\Text(130,110)[b]{
{\color{Blue}$\phi_1,\phi_2$}}
\DashArrowArc(65,100)(35,0,180){5}\Text(65,147)[]{
{\color{Blue}$\tilde{t}_1,\tilde{t}_2,
\tilde{t}_1,\tilde{t}^*_1$}}
\DashArrowArc(65,100)(35,180,360){5}\Text(65,53)[]{
{\color{Blue}$\tilde{t}_1,\tilde{t}_2,
\tilde{t}_2,\tilde{t}^*_2$}}


\DashArrowLine(200,100)(240,100){5}\Text(210,110)[br]{{\color{Red}$a$}}
\DashArrowLine(240,100)(280,100){5}\Text(280,110)[b]{
{\color{Blue}$\phi_1,\phi_2$}}
\Text(240,100.5)[]{\boldmath $\times$}\Text(240,110)[b]{$T_{\color{Red}a}$}


\DashArrowArc(380,120)(30,0,360){5}\Text(380,160)[b]{
{\color{Blue}$\tilde{t}_1,\tilde{t}_2$} }
\DashArrowLine(380,50)(380,90){5}\Text(385,60)[l]{{\color{Red}$a$}}

\end{picture}
\end{center}
\vspace{-2cm}
\caption{One-loop contributions to scalar-pseudoscalar mixing in the
MSSM.}\label{f2}
\end{figure}
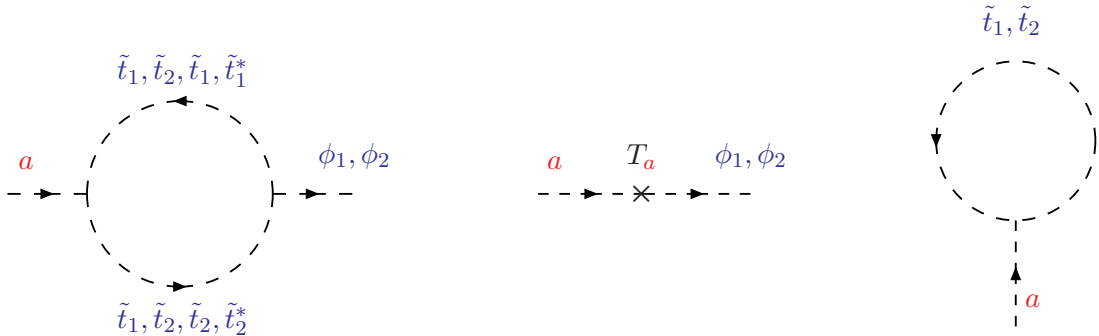

In the constrained MSSM with universal soft supersymmetry-breaking masses
$m_0$ for sfermions and $m_{1/2}$ for gauginos, as well as trilinear
supersymmetry-breaking parameters $A_i$, there are two independent
CP-violating phases. Without loss of generality, these may be taken as the
phase of the gluino mass $m_{\tilde g}$ and $A_t$. 
As illustrated in Fig.~\ref{f2}, CP-violating
scalar-pseudoscalar transitions ${\cal M}^2_{SP} \sim \widehat{\Pi}^{HA}(0)$
in the MSSM are predominantly induced by stop squarks. The qualitative 
behaviours of important CP-violating terms are given by
\begin{equation}
  \label{MSP}
{\cal  M}^2_{SP}\ \sim\ \frac{m^4_t}{v^2}\, 
\frac{{\rm Im}\, (\mu A_t)}{32\pi^2\, M^2_{\rm SUSY}}\ 
\bigg( 1,\ \frac{|A_t|^2}{M^2_{\rm SUSY}}\,,\ 
\frac{|\mu|^2}{\tan\beta\,M^2_{\rm SUSY}}\,,\ 
\frac{2{\rm Re}\, (\mu A_t)}{M^2_{\rm SUSY}}\,\bigg)\, ,
\label{SPmixing}
\end{equation}
where $\mu$ is the supersymmetric parameter characterizing the mixing
of the two Higgs superfields, and $M^2_{\rm SUSY}$ specifies the
common soft supersymmetry-breaking scale defined by the arithmetic
average of the squared stop masses. We see from~(\ref{MSP}) that
${\cal M}^2_{SP}$ could in principle be as large as $M^2_Z$~\cite{APLB}.

This is possible if the phase of $A_t$ is large, which cannot be
excluded {\it a priori}. There are important constraints on the
CP-violating phases in the MSSM coming, in particular, from
constraints on the electric dipole moments of the electron, neutron
and ${}^{199}$Hg~\cite{FOPR}.  However, cancellations are possible
between different supersymmetric diagrams, and between different
CP-violating operators~\cite{EF,IN}.  Moreover, the constraints apply
directly only to the first and possibly second generation of matter
fermions, and so may be more relaxed for the third-generation coupling
$A_t$~\cite{CKP}, if one relaxes the assumption of universality
between the different generations.

Analogously, in heavy Majorana-neutrino models, scalar-pseudoscalar
mixings are induced after integrating out the heavy Majorana neutrinos,
and may also be non-negligible. However, we do not discuss such models in 
any further detail here.

\begin{figure}[t]
\begin{center}
\begin{picture}(320,150)(0,0)
\SetWidth{0.7}
 
\ArrowLine(0,70)(30,70)\ArrowLine(30,70)(60,70)
\ArrowLine(60,70)(90,70)\ArrowLine(90,70)(120,70)
\DashArrowArc(60,70)(30,90,180){3}\DashArrowArc(60,70)(30,0,90){3}
\DashArrowLine(60,130)(60,100){3}
\Text(60,70)[]{\boldmath $\times$}
\Text(-4,65)[lt]{$b_L$}\Text(130,65)[rt]{$b_R$}
\Text(60,62)[t]{{\color{BrickRed}$\tilde{g}$}}\Text(60,125)[bl]{
{\color{OliveGreen}$\phi^{0\ast}_{1,2}$} }
\Text(34,100)[r]{{\color{Blue}$\tilde{b}^*_L$}}
\Text(86,100)[l]{{\color{Blue}$\tilde{b}^*_R$}}

\ArrowLine(200,70)(230,70)\ArrowLine(230,70)(260,70)
\ArrowLine(260,70)(290,70)\ArrowLine(290,70)(320,70)
\DashArrowArc(260,70)(30,90,180){3}\DashArrowArc(260,70)(30,0,90){3}
\DashArrowLine(260,130)(260,100){3}
\Text(260,70)[]{\boldmath $\times$}
\Text(196,65)[lt]{$b_L$}\Text(330,65)[rt]{$b_R$}
\Text(245,65)[t]{{\color{BrickRed}$\tilde{h}^-_2$} }
\Text(285,65)[t]{{\color{BrickRed}$\tilde{h}^-_1$} }
\Text(265,125)[bl]{{\color{OliveGreen}$\phi^{0\ast}_{1,2}$}}
\Text(234,100)[r]{{\color{Blue}$\tilde{t}^*_R$}}
\Text(286,100)[l]{{\color{Blue}$\tilde{t}^*_L$}}

\end{picture}
\begin{picture}(320,100)(0,0)
\SetWidth{1.2}
 
\ArrowLine(0,70)(30,70)\ArrowLine(30,70)(60,70)
\ArrowLine(60,70)(90,70)\ArrowLine(90,70)(120,70)
\DashArrowArc(60,70)(30,90,180){3}\DashArrowArc(60,70)(30,0,90){3}
\DashArrowLine(60,130)(60,100){3}
\Text(60,70)[]{\boldmath $\times$}
\Text(-4,65)[lt]{$t_L$}\Text(130,65)[rt]{$t_R$}
\Text(60,62)[t]{{\color{BrickRed}$\tilde{g}$}}\Text(60,125)[bl]{
{\color{OliveGreen}$\phi^0_{1,2}$} }
\Text(34,100)[r]{{\color{Blue}$\tilde{t}^*_L$}}
\Text(86,100)[l]{{\color{Blue}$\tilde{t}^*_R$}}

\ArrowLine(200,70)(230,70)\ArrowLine(230,70)(260,70)
\ArrowLine(260,70)(290,70)\ArrowLine(290,70)(320,70)
\DashArrowArc(260,70)(30,90,180){3}\DashArrowArc(260,70)(30,0,90){3}
\DashArrowLine(260,130)(260,100){3}
\Text(260,70)[]{\boldmath $\times$}
\Text(196,65)[lt]{$t_L$}\Text(330,65)[rt]{$t_R$}
\Text(245,65)[t]{{\color{BrickRed}$\tilde{h}^+_1$} }
\Text(285,65)[t]{{\color{BrickRed}$\tilde{h}^+_2$} }
\Text(265,125)[bl]{{\color{OliveGreen}$\phi^0_{1,2}$}}
\Text(234,100)[r]{{\color{Blue}$\tilde{b}^*_R$}}
\Text(286,100)[l]{{\color{Blue}$\tilde{b}^*_L$}}

\end{picture}
\end{center}
\vspace{-2cm}
\caption{CP-violating vertex effects on the bottom and
  top Yukawa couplings.}\label{f2v}
\end{figure}
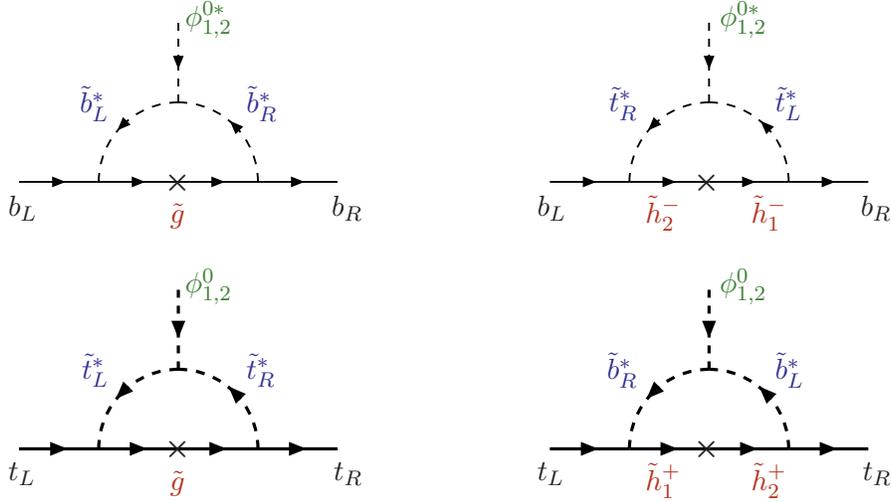

\subsection{CP-Violating Vertex Effects}

In addition to the CP-violating self-energy effects, CP-violating
vertex effects involving gluinos, higgsinos  and squarks of the  third
generation   \cite{CEPW}    may  drastically   modify    the  top- and
bottom-quark Yukawa couplings  $h_t$ and $h_b$. These effects
could  play   a significant  r\^ole not  only   in the effective Higgs
potential~\cite{CEPW} at the two-loop level, but may  also
affect directly the production of reconstructed  polarized top and
bottom quarks. We therefore generalize the discussion of CP-conserving
vertex effects in the previous subsection of this report to include 
CP-violating vertex corrections. The Feynman diagrams shown
in Fig.~\ref{f2v} induce the following CP-violating effective Lagrangian
for the couplings:
\begin{equation}
  \label{Lvert}
-{\cal L}^{\rm eff} \ =\ \Big[\, (h_b + \delta h_b)\, \phi^{0*}_1\: +\: 
\Delta h_b \phi^{0*}_2\, \Big]\,\bar{b}_R b_L\ +\ 
\Big[\, (h_t + \delta h_t)\, \phi^0_2\: +\: 
\Delta h_t \phi^0_1\, \Big]\,\bar{t}_R t_L\ +\ {\rm h.c.},
\end{equation}
where 
\begin{eqnarray}
  \label{deltahbt}
\frac{\delta h_b}{h_b} &\sim& -\, \frac{2\alpha_s}{3\pi}\,
\frac{m^*_{\tilde{g}} A_b}{{\rm max}\, \big( M^2_{\rm SUSY},\ 
|m_{\tilde{g}}|^2\big)}\ -\  \frac{|h_t|^2}{16\pi^2}\,
\frac{|\mu |^2}{{\rm max}\, \big( M^2_{\rm SUSY},\ 
|\mu |^2\big)}\ ,\nonumber\\
\frac{\Delta h_b}{h_b} &\sim&  \frac{2\alpha_s}{3\pi}\,
\frac{m^*_{\tilde{g}} \mu^*}{{\rm max}\, \big( M^2_{\rm SUSY},\ 
|m_{\tilde{g}}|^2\big)}\ +\  \frac{|h_t|^2}{16\pi^2}\,
\frac{A^*_t \mu^*}{{\rm max}\, \big( M^2_{\rm SUSY},\ 
|\mu |^2\big)}\ ,\nonumber\\
\frac{\delta h_t}{h_t} &\sim& -\, \frac{2\alpha_s}{3\pi}\,
\frac{m^*_{\tilde{g}} A_t}{{\rm max}\, \big( M^2_{\rm SUSY},\ 
|m_{\tilde{g}}|^2\big)}\ -\  \frac{|h_b|^2}{16\pi^2}\,
\frac{|\mu |^2}{{\rm max}\, \big( M^2_{\rm SUSY},\ 
|\mu |^2\big)}\ ,\nonumber\\
\frac{\Delta h_t}{h_t} &\sim&  \frac{2\alpha_s}{3\pi}\,
\frac{m^*_{\tilde{g}} \mu^*}{{\rm max}\, \big( M^2_{\rm SUSY},\ 
|m_{\tilde{g}}|^2\big)}\ +\  \frac{|h_b|^2}{16\pi^2}\,
\frac{A^*_b \mu^*}{{\rm max}\, \big( M^2_{\rm SUSY},\ 
|\mu |^2\big)}\ ,
\end{eqnarray}
and
\begin{eqnarray}
  \label{htb}
h_b &=& \frac{ g_w m_b}{\sqrt{2}\, M_W\, \cos\beta\, [\, 1\:
+\: \delta h_b/h_b\: +\: (\Delta h_b/h_b)\, \tan\beta\, ]}\ ,\nonumber\\ 
h_t &=& \frac{ g_w m_t}{\sqrt{2}\, M_W\, \sin\beta\, [\, 1\:
+\: \delta h_t/h_t\: +\: (\Delta h_t/h_t)\, \cot\beta\, ]}\ .
\end{eqnarray}
We see from these last relations that the modification of
the
Higgs-bottom  Yukawa   coupling  is sizeable  for   large   values  of
$\tan\beta$, whilst the corresponding corrections  to the Higgs-top Yukawa
coupling are less relevant.

\subsection{CP Asymmetries}

\begin{figure}[bht]
\begin{center}
\begin{picture}(300,100)(0,0)
\SetWidth{0.7}

\Line(15,10)(55,90)
\Line(55,90)(285,90)\Line(285,90)(245,10)\Line(245,10)(15,10)
\LongArrow(15,50)(145,50)\Text(0,50)[r]{${\color{Blue}\mu^-}
({\bf {\rm p_-, s_-}})$}
\LongArrow(285,50)(155,50)\Text(300,50)[l]{${\color{Blue}\mu^+}
({\bf {\rm p_+, s_+}})$}
\Vertex(150,50){3}
\SetWidth{2}
\LongArrow(75,50)(75,75)\LongArrow(75,50)(100,50)
\Text(100,35)[]{${\color{Blue}P^-_L}$}
\LongArrow(235,50)(215,70)\LongArrow(235,50)(205,50)
\Text(205,35)[]{${\color{Blue}P^+_L}$}
\SetWidth{1.0}
\Text(209,109)[]{${\color{Blue}f}$}\LongArrow(150,50)(200,100)
\LongArrow(150,50)(100,0)\Text(90,-10)[]{${\color{Blue}\bar{f}}$}
\Text(100,65)[]{${\color{Red}\phi_{_-}}$}\ArrowArc(75,50)(20,60,90)
\Text(232,85)[]{${\color{Red}\phi_{_+}}$}\ArrowArc(235,50)(20,60,135)
\Text(62,79)[]{${\color{Blue}P^-_T}$}\DashLine(75,50)(95,90){3}
\Text(190,62)[]{{\color{Blue}$\theta$}}\ArrowArc(150,50)(30,0,45)
\Text(203,79)[]{${\color{Blue}P^+_T}$}\DashLine(235,50)(255,90){3}

\end{picture}
\end{center}       
\caption{CP asymmetries with polarized muons.}
\label{f3}
\end{figure}
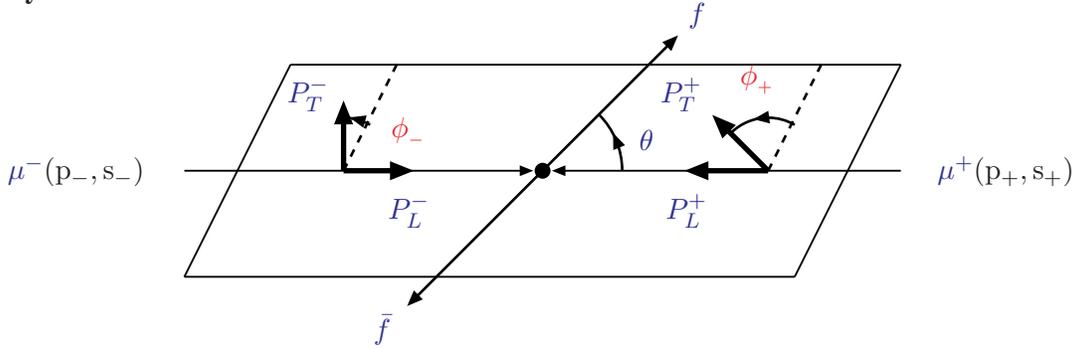

The possibility of muon polarization at a \mm\ collider could play an 
essential role in
unravelling  the CP nature  of  the Higgs boson(s) and/or in probing CP
violation in  the Higgs sector.  We display in Fig.~\ref{f3} a general
configuration  of the polarizations of the initial  muons. There  are
several CP-violating observables that can be constructed  using muon
polarization vectors and/or the  three-momenta and spins of the  final
particles.  For our illustrations, however, we concentrate on
the following two representative CP-odd observables \cite{GG,ANPB}:
\begin{eqnarray}
  \label{AtCP}
{\cal A}^t_{\rm CP} &=& \frac{\sigma (\mu^- (s_x) \mu^+ (s_y)\to
f\bar{f})\: -\: \sigma (\mu^- (s_x) \mu^+ (-s_y)\to
f\bar{f}) }{ \sigma (\mu^- (s_x) \mu^+ (s_y)\to f\bar{f} )\: 
+\: \sigma (\mu^- (s_x) \mu^+ (-s_y)\to f\bar{f} )}\ ,\\
  \label{AlCP}
{\cal A}^l_{\rm CP} &=& \frac{\sigma (\mu^- (s_z) \mu^+ (-s_z)\to
f\bar{f})\: -\: \sigma (\mu^- (-s_z) \mu^+ (s_z)\to
f\bar{f}) }{ \sigma (\mu^- (s_z) \mu^+ (-s_z)\to f\bar{f} )\: 
+\: \sigma (\mu^- (-s_z) \mu^+ (s_z)\to f\bar{f} )}\ ,
\end{eqnarray}
where $s_{x,y,z}$ are the $x,y,z$-projections of the spin {\bf $s$} of
the muon.  Note that we define the positive $z$ axis as the direction
of the $\mu^-$ beam, and the $y$ axis perpendicular to the earth
surface pointing upwards to the sky.  The CP-violating observable
${\cal A}^t_{\rm CP}$ is even under naive CP`T' transformations,
whereas ${\cal A}^l_{\rm CP}$ is odd.  To leading order, ${\cal
A}^t_{\rm CP}$ is generated by dispersive terms, whilst ${\cal
A}^l_{\rm CP}$ requires non-vanishing absorptive contributions.

\begin{figure}[p]
 \begin{center}
    \epsfig{figure=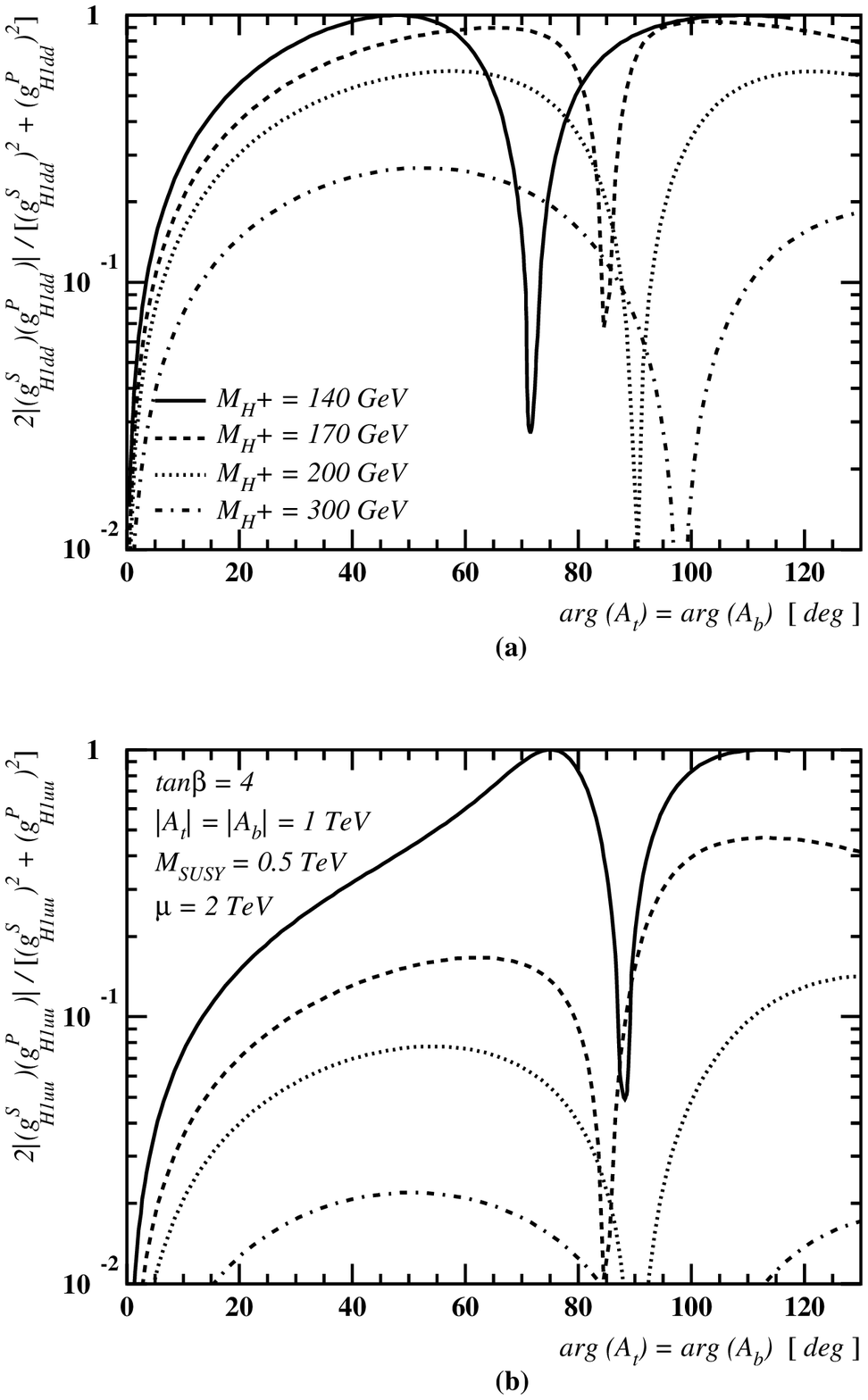,height=18cm}
 \end{center}
\vspace*{-1cm}
\caption{Numerical estimates of (a) $2|(g^S_{H_1dd})\, (g^P_{H_1dd})|/
  [(g^S_{H_1dd})^2 + (g^P_{H_1dd})^2]$ and (b) $2|(g^S_{H_1uu})\,
  (g^P_{H_1uu})|/[(g^S_{H_1uu})^2 + (g^P_{H_1uu})^2]$ as functions of
  ${\rm arg} (A_t )$.}\label{f4}
\end{figure}

\begin{figure}[p]
 \begin{center}
    \epsfig{figure=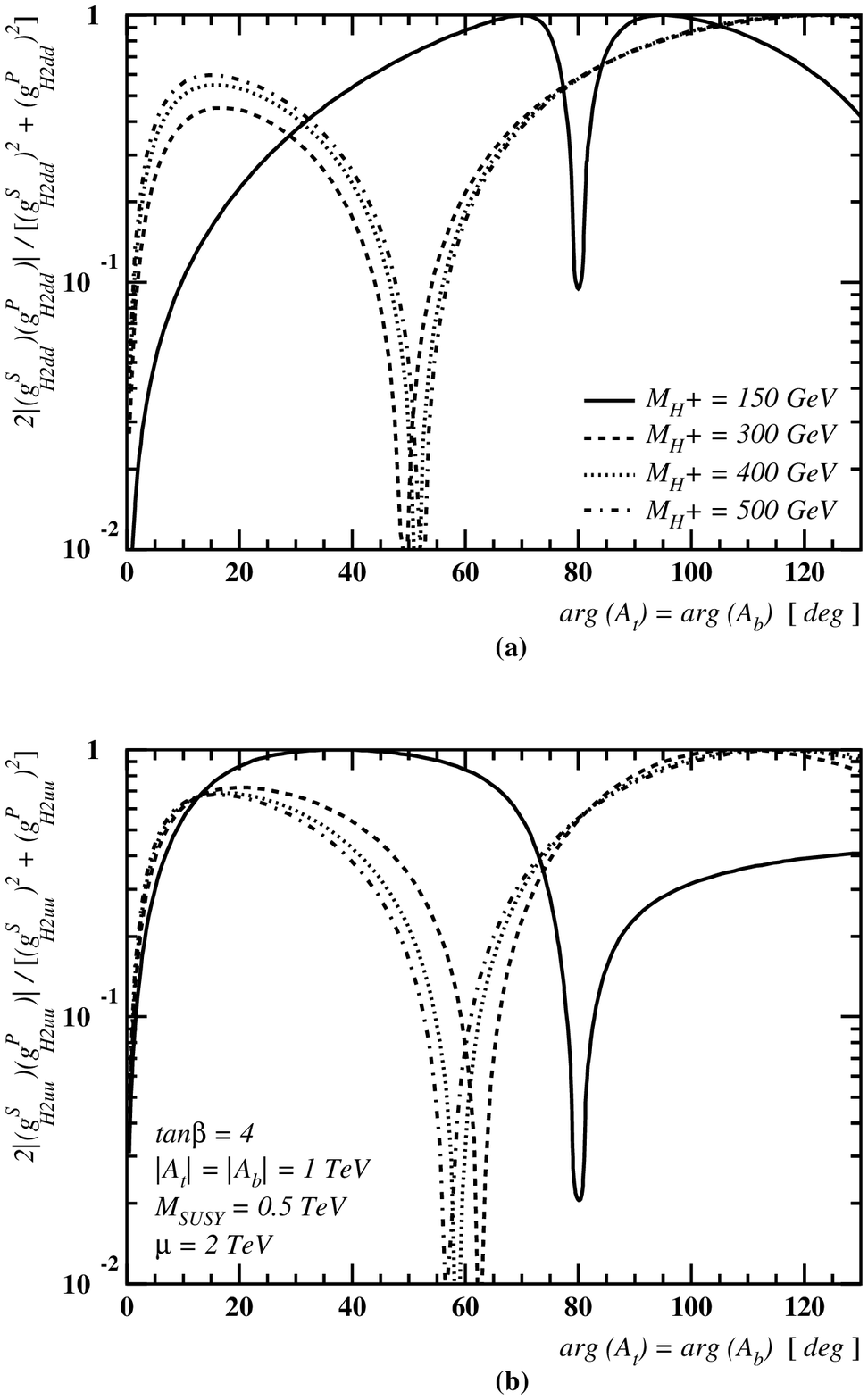,height=18cm}
 \end{center}
\vspace*{-1cm}
\caption{Numerical estimates of (a) $2|(g^S_{H_2dd})\, (g^P_{H_2dd})|/
  [(g^S_{H_2dd})^2 + (g^P_{H_2dd})^2]$ and (b) $2|(g^S_{H_2uu})\,
  (g^P_{H_2uu})|/[(g^S_{H_2uu})^2 + (g^P_{H_2uu})^2]$ as functions of
  ${\rm arg} (A_t )$.}\label{f5}
\end{figure}

The interactions of Higgs bosons $H_i$ with mixed CP 
to fermions $f$ are given by
\begin{equation}
  \label{Lff}
{\cal L}_{\rm int}\ =\ -\sum_{i=1}^3\,H_i\, \frac{g_w m_f}{2M_W}\, 
\bar{f}\Big( \, g^S_{H_iff}\: +\: ig^P_{H_iff} \, \Big) f\, .
\end{equation}
In  the MSSM,     the   reduced scalar  and pseudoscalar     couplings
$g^S_{H_iff}$   and   $g^P_{H_iff}$ receive  contributions  from  both
self-energy and  vertex  corrections similar to those discussed 
previously in the CP-conserving case, and their  analytic  forms have 
been derived in~\cite{CEPW}. Neglecting the $\gamma,\
Z$ background, as appropriate at energies close to any Higgs-boson
resonance $H_i: i = 1,2,3$, the
CP-violating observable ${\cal A}^t_{\rm CP}$ reads
\begin{equation}
  \label{AtCPMSSM}
{\cal A}^t_{\rm CP}\ =\ \frac{2 g^S_{H_i\mu\mu}\, g^P_{H_i\mu\mu}}{
(g^S_{H_i\mu\mu})^2\: +\: (g^P_{H_i\mu\mu})^2}\ .
\end{equation}
We display in Fig.~\ref{f4}(a) the dependence of ${\cal
A}^t_{\rm
CP}$ on ${\rm arg}\, (A_t)$ for the lightest  Higgs boson $H_1$, and
Fig.~\ref{f4}(b)  shows numerical estimates of ${\cal A}^t_{\rm
CP}$ for  transversely-polarized up-type  fermions.  Correspondingly,
numerical estimates related  to the  next-to-lightest $H_2$ boson  are
exhibited in Fig.~\ref{f5}. 

\begin{figure}[p]
 \begin{center}
    \epsfig{figure=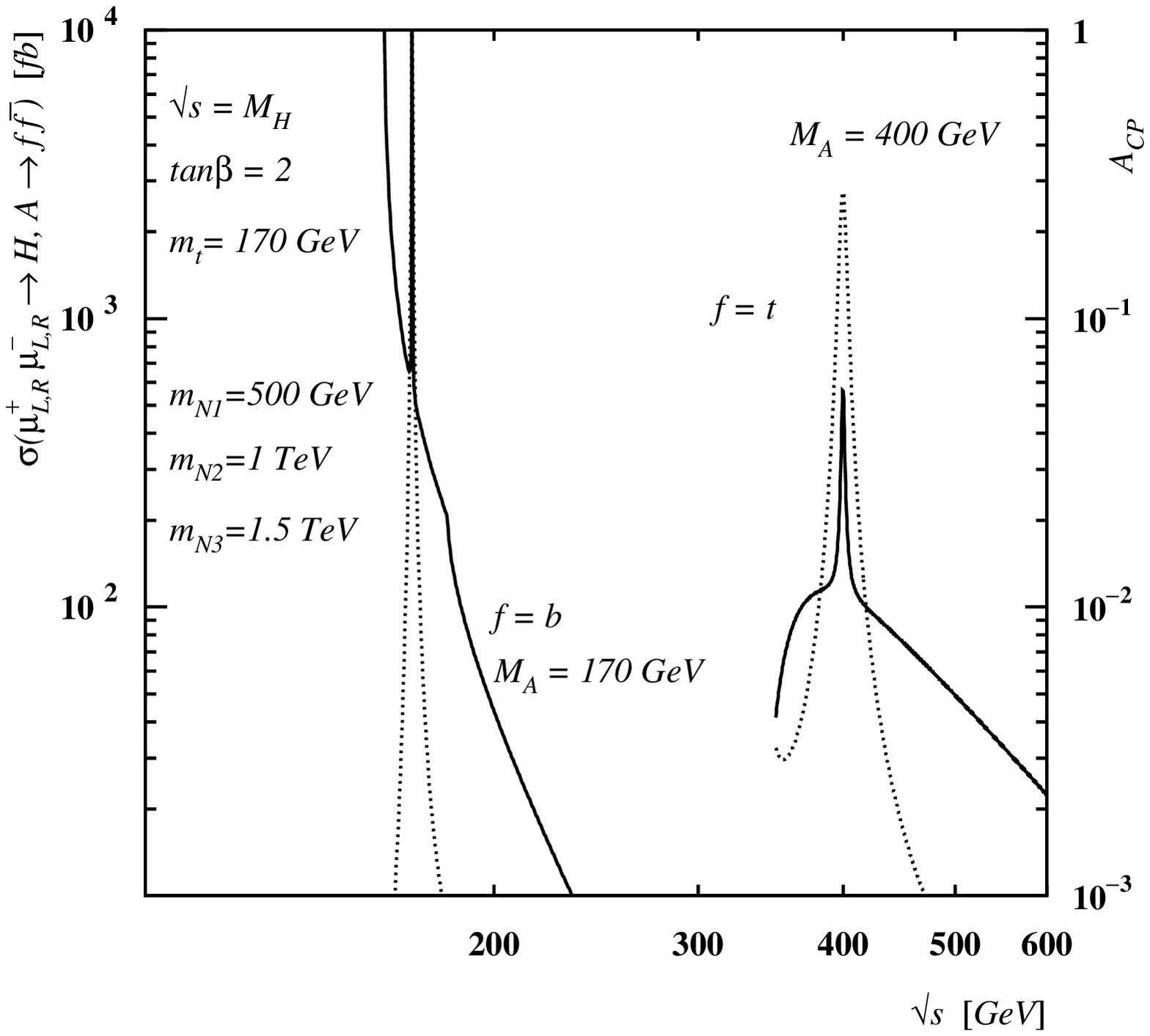,height=18cm}
 \end{center}
\vspace*{-5cm}
\caption{Numerical values for the cross section (solid lines) and 
${\cal A}^l_{\rm CP}$ (dashed lines) as functions of the cms energy 
$\sqrt{s}$, for an $E_6$-inspired model with heavy Majorana
neutrinos~\cite{ANPB}.}\label{f6}
\end{figure}

\begin{figure}[p]
\begin{center}
    \epsfig{figure=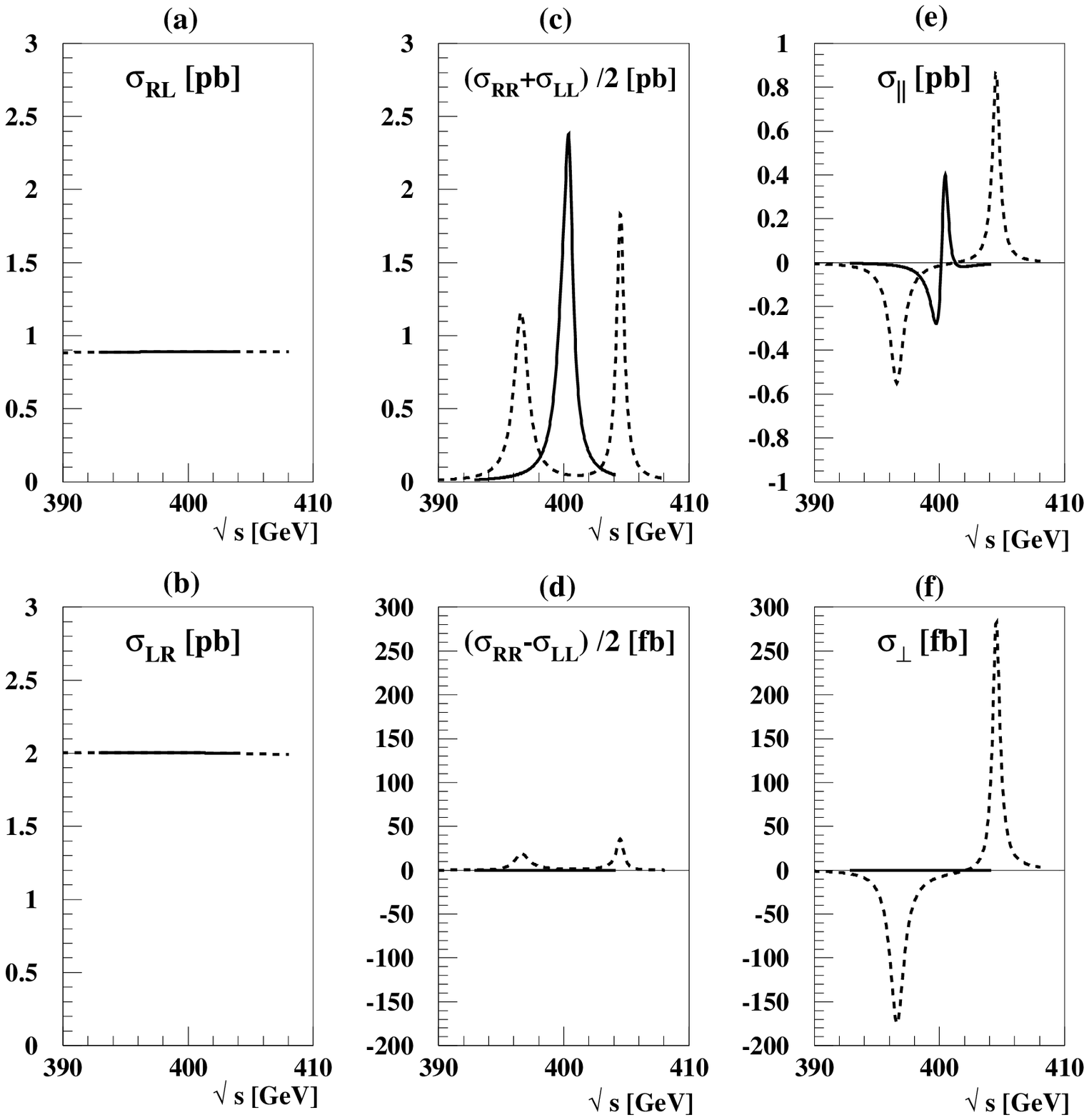,height=16cm}
\end{center}
\caption{
The dependence of polarized-muon cross sections on the centre-of-mass
energy $\sqrt{s}$, as estimated in~\cite{SYC} for the MSSM with explicit
CP violation.  The selected parameters are: $\tan\beta =3$, $M_{H^+}
\approx 0.4$~TeV, $M_{\rm SUSY} = 0.5$~TeV, $|A_{t,b}| = |\mu| = 1$~TeV
and ${\rm arg} (\mu A_t) = 0$ (solid lines) and $90^\circ$ (dashed
lines).}
\label{f7}
\end{figure}

The   CP-violating  observable   ${\cal   A}^l_{\rm  CP}$ defined   by 
(\ref{AlCP}) may be approximated by~\cite{ANPB}:
\begin{equation}
  \label{AlCP1}
{\cal   A}^l_{\rm  CP} \ \approx\ 
\frac{2 {\rm Re}\,(\widehat{\Pi}^{HA})\  {\rm Im}\, (\widehat{\Pi}^{AA} - 
\widehat{\Pi}^{HH} )}{(M^2_H - M^2_A)^2\: +\: 
({\rm Im}\,\widehat{\Pi}^{AA})^2 + ({\rm Im}\,\widehat{\Pi}^{HH})^2}\ .
\end{equation}
The expression~(\ref{AlCP1}) is derived under the assumption that only
one CP-even Higgs boson $H$ mixes actively with a CP-odd scalar $A$,
after integrating out heavy degrees of freedom which amounts to a 
vanishing or rather suppressed ${\rm Im}\, \widehat{\Pi}^{HA}$ for energies
below the TeV scale.  Such a scenario has been studied in~\cite{ANPB},
within the context of a heavy-Majorana-neutrino model.  We display in
Fig.~\ref{f6} numerical values for two scenarios with $M_A = 170$ and
400~GeV.  In agreement with our earlier discussion, we observe that
${\cal A}^l_{\rm CP}$ may become of order unity if $M_H - M_A \sim
\Gamma_H,\ \Gamma_A$. As was discussed  in \cite{SYC} and is displayed
in Fig.~\ref{f7}(c), (d) and (f), analogous features may be found in
the MSSM with explicit radiative CP violation in the Higgs sector,
where ${\cal A}^t_{\rm CP} = \sigma_{\perp}/(\sigma_{LL} +
\sigma_{RR})$ and ${\cal A}^l_{\rm CP} = (\sigma_{LL} -
\sigma_{RR})/(\sigma_{LL} +
\sigma_{RR})$.

This pilot study indicates that the option of polarization for the $\mu^+$
and $\mu^-$ beams may be very valuable for determining the CP nature of a
Higgs boson and/or for analyzing a two-Higgs-boson-mixing system.  
However, the effective degree of polarization provided naturally in a \mm
collider is currently not expected to be much larger than $P\sim 0.4$ for
each beam~\cite{GGP}.  Thus, the actual CP asymmetries must be reduced by
a factor $P^2\sim 1/10$ compared to the above predictions. Nevertheless,
there may well be large observable effects. A complete evaluation of this
opportunity requires further studies of both theoretical and experimental
aspects, including background analyses originating from $\gamma,\
Z$-exchange graphs as well as the effects of polarization dilution.

\subsection{Heavy Neutral Higgs-Boson Masses in the MSSM 
            with Explicit CP Violation}

\begin{figure}[p]
\begin{center}
   \leavevmode
   \epsfysize=19.5cm
   \epsfig{figure=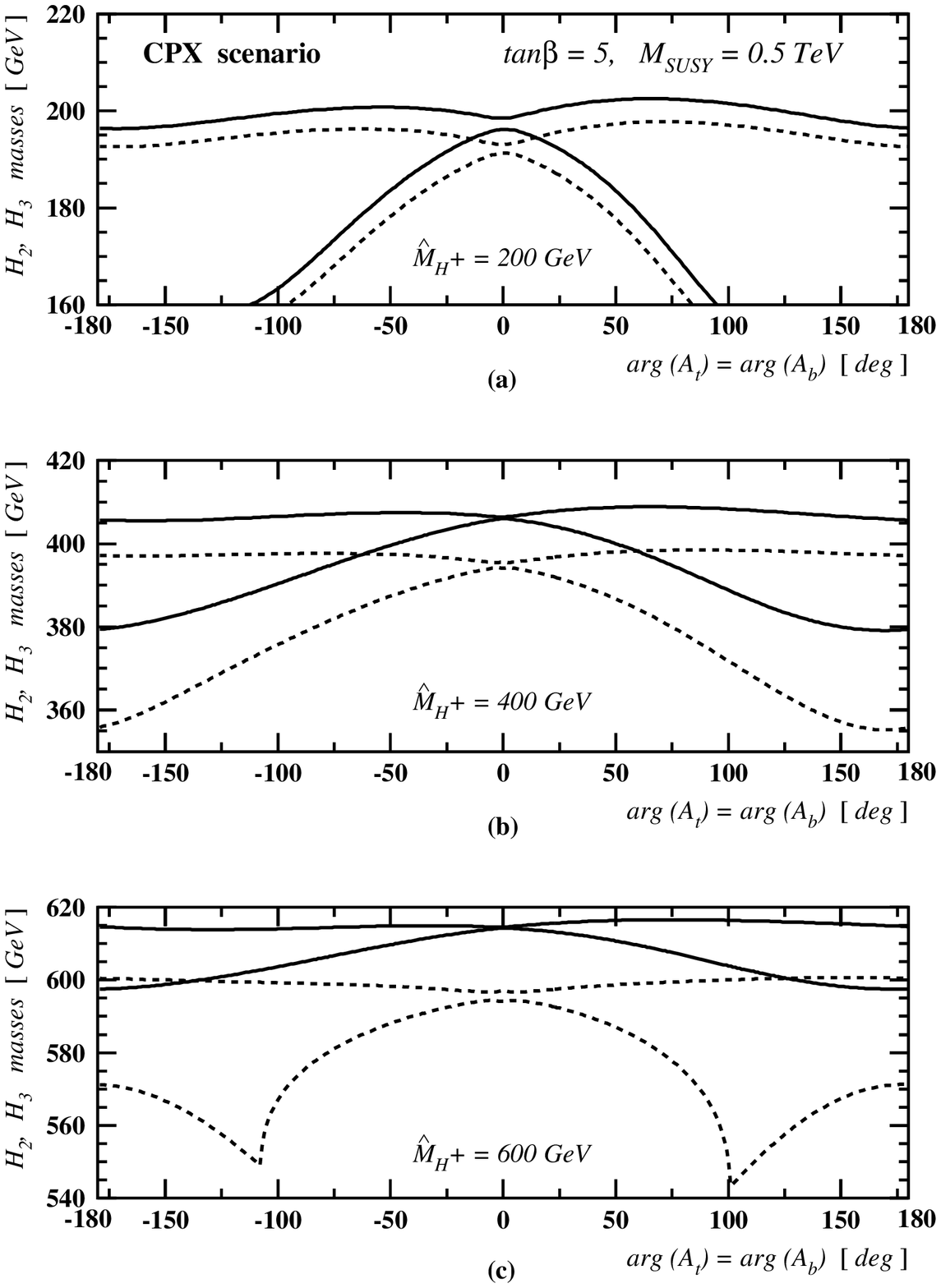, height=16cm}
\end{center}
\caption{Numerical values for the mass splitting between the second and
third neutral Higgs bosons $H_{2,3}$ of the MSSM, as functions of 
Arg($A_t$), the
CP-violating phase of $A_t$, for the indicated values of the other MSSM
parameters. There are significant differences between the mass 
differences calculated in the effective-potential approximation (solid 
lines) and using the pole masses. We note that the mass splitting may be 
increased by a large factor for large values of Arg($A_t$)~\cite{CEPW}.}
\label{fig:pole3}
\end{figure}

We demonstrated earlier how resonant Higgs scalar-pseudoscalar transitions
may lead to enhanced CP asymmetries at a \mm\ collider. We now explore 
an appealing theoretical framework for such studies, namely the MSSM with
explicit radiative CP violation. In this case, the lightest neutral Higgs
boson $H_1$ already offers interesting physics prospects.  However,
further opportunities are offered by the two heavier neutral Higgs bosons
$H_{2,3}$, which are in general largely mixtures of the heavier `CP-even'
Higgs boson $H$ and the `CP-odd' pseudoscalar $A$, which can naturally
have a mass splitting comparable to their widths \cite{APLB,CEPW}. 
This mass splitting will in general be increased by scalar-pseudoscalar 
mixing (\ref{SPmixing}), as seen
in Fig.~\ref{fig:pole3}. This figure contrasts recent results found 
for pole masses of $H_{2,3}$ \cite{cepw3} using the code {\tt cph$+$.f} 
\cite{cphplus} with previous results found in an
effective-potential approach \cite{CEPW}. 
We see that these calculations are
significantly different, in particular for larger values of the charged
Higgs-boson mass, $m_{H^+}$, and in the vicinity of the stop threshold.

The results are plotted as functions of Arg($A_t$), the CP-violating phase
of $A_t$. We see that the difference in masses between the two heavier
neutral Higgs bosons may easily be increased by a large factor if
Arg($A_t$) is large. The detailed study of CP-violating observables as a
function of the centre-of-mass energy across the double $H_{2,3}$ peak is
left for another occasion, but it is clear that the beam polarization 
discussed earlier would be a valuable tool, as well as studies of the 
polarization states of $H_{2,3}$ decay products.

\section{Precise Determination of the Higgs--Chargino Couplings 
            in Chargino-Pair Production}

\begin{figure}[ht]
\begin{center}
\unitlength1cm
\begin{picture}(14,4)
\put(0.9,-.6){\includegraphics{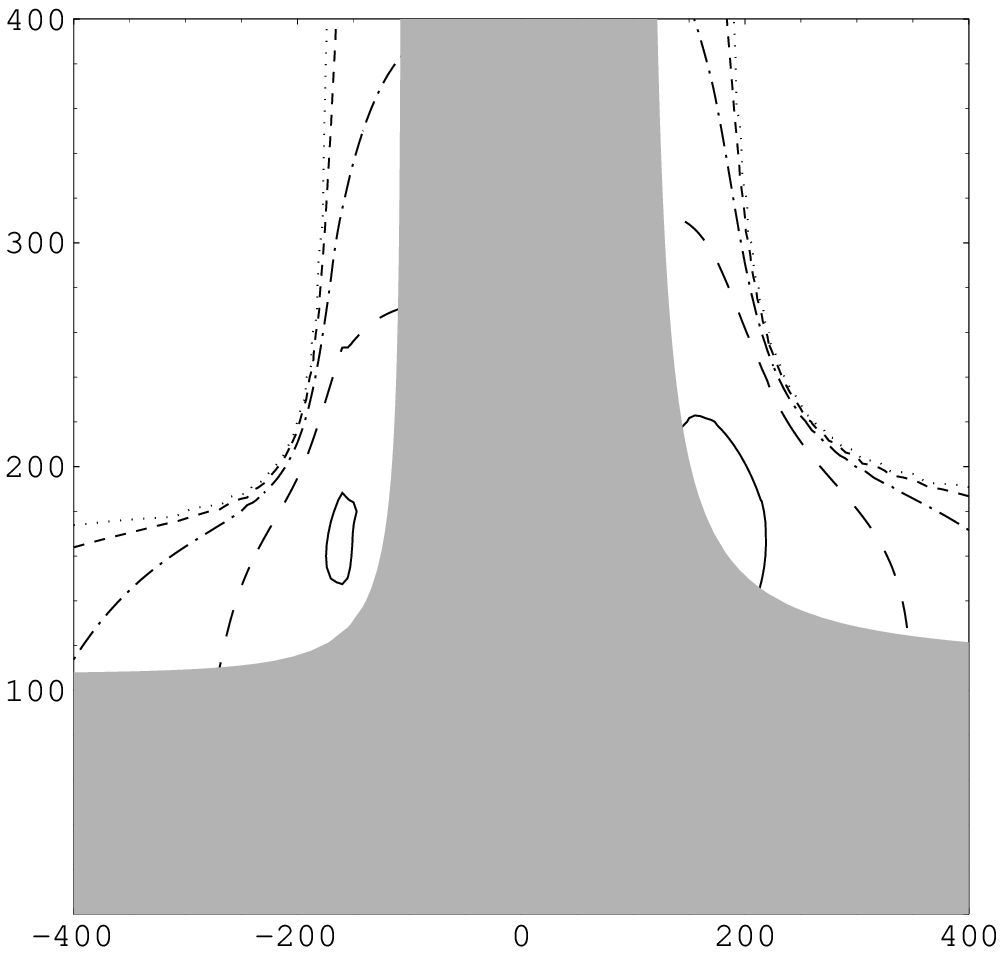}}
\put(7.9,-.6){\includegraphics{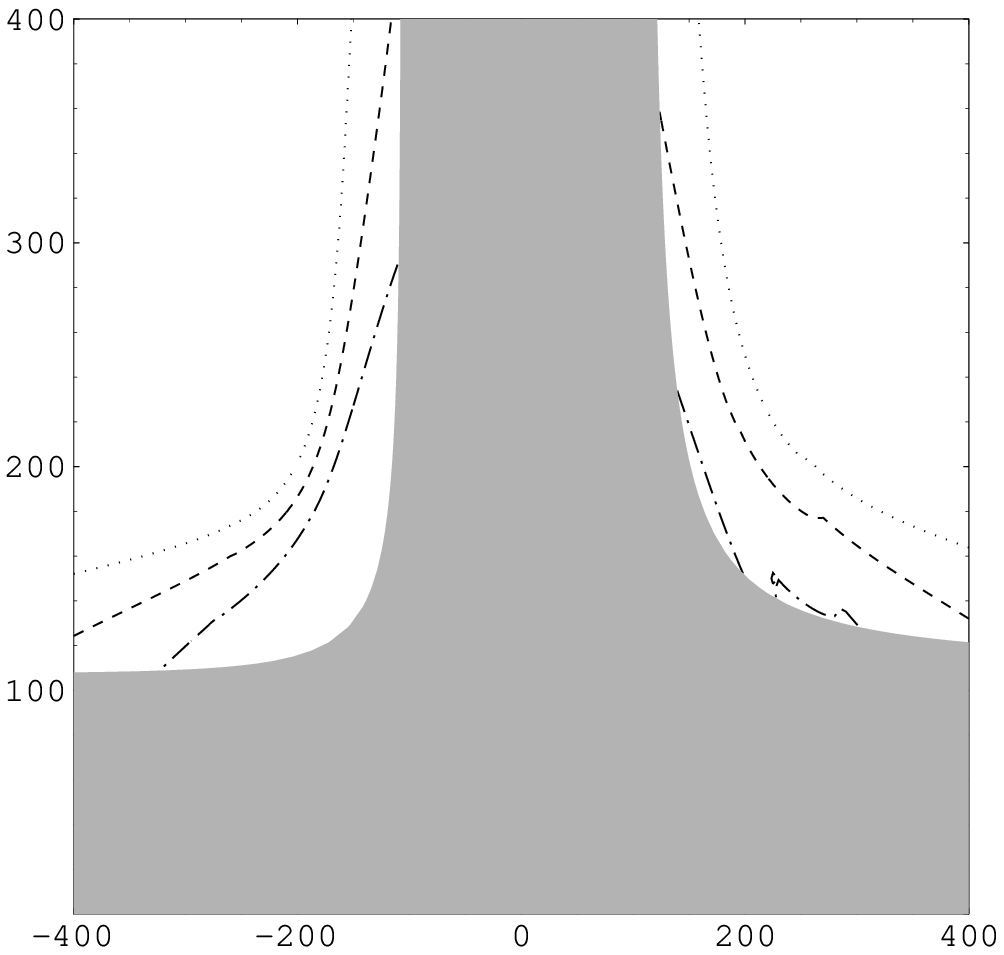}}
\put(1.5,3.4){$ \scriptstyle (a)$}
\put(8.5,3.4){$ \scriptstyle (b)$}
\put(2.3,4){$ \scriptstyle BR(A \rightarrow 
                  \tilde{\chi}^+_1 \tilde{\chi}^-_1)$}
\put(-.2,3.4){$ \scriptstyle M_2/GeV$}
\put(4.4,-.8){$ \scriptstyle \mu/GeV$}
\put(9.3,4){$ \scriptstyle BR(H \rightarrow 
                  \tilde{\chi}^+_1 \tilde{\chi}^-_1)$}
\put(6.8,3.4){$ \scriptstyle M_2/GeV$}
\put(11.4,-.8){$ \scriptstyle \mu/GeV$}
\end{picture}
\end{center}
\vspace{2mm}
\caption{
        Branching ratios of the heavy Higgs bosons
        $A$ and $H$ 
        into light-chargino pairs for $m_A=350$~GeV and $\tan\beta=5$,
        computed with the program {\tt HDECAY}~\cite{hdecay}.
        The contour lines correspond to 
        0.1 (dotted), 0.2 (dashed), 0.3 (dash-dotted), 
        0.4 (large dashed) and 0.5 (continuous).
        The gray area is the experimentally excluded region.}
\label{fig:Wien1}
\end{figure}

We now give one example of the interesting CP-conserving physics
accessible at a second \mm $(H,A)$ factory:
chargino-pair production at such a \mm\ collider offers an outstanding
possibility for the precise determination of the Higgs-chargino
couplings. 
Within the framework of the MSSM, 
decays of the heavy neutral Higgs bosons $H$, $A$
into two light charginos can be
observed with significant branching ratios in certain
parameter regions. Fig.~\ref{fig:Wien1} shows branching ratios up to
50 \% in mixed scenarios with 
$|\mu| \sim M_2$ for $\tan\beta=5$ and $m_A=350$~GeV,
assuming CP symmetry~\footnote{There are interesting additional
physics opportunities in chargino production and decay in the MSSM with
CP violation, extending the analysis of the previous section, which we 
leave for a future occasion.}.
We study the pair production of light charginos
with mass $m_{\tilde{\chi}^\pm_1}=155$~GeV 
in such a mixed scenario with $M_2 = -\mu = 188$~GeV,
and in a scenario with a gaugino-dominated light chargino:
$M_2 = 155$~GeV, $\mu = -400$~GeV for comparison.

\begin{figure}[htb]
\begin{center}
\unitlength1cm
\begin{picture}(12,3)
\put(-0.5,-7.6){\includegraphics{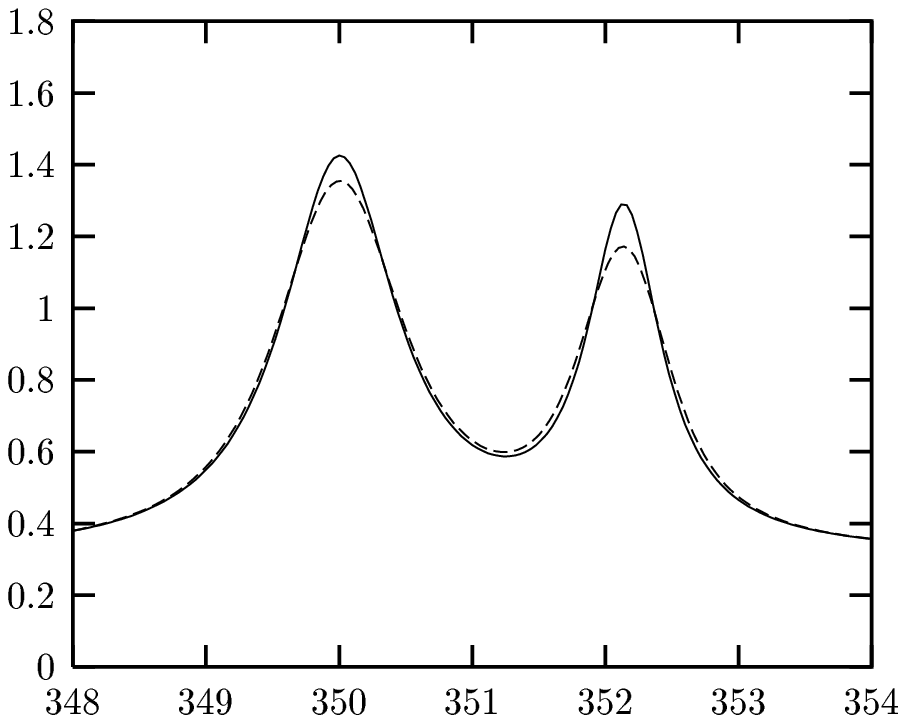}}
\put(6.5,-7.6){\includegraphics{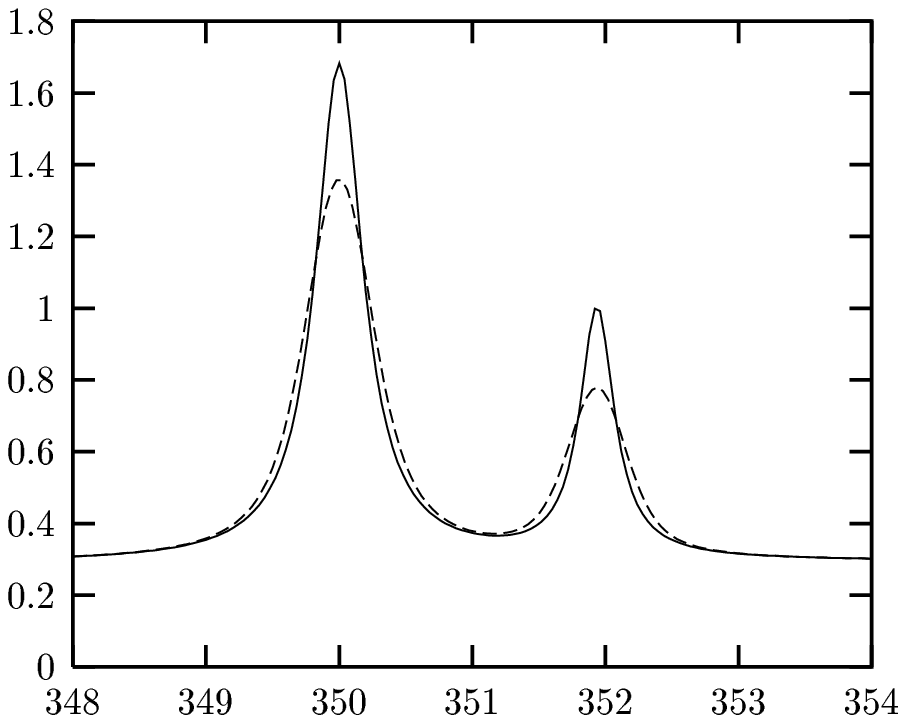}}
\put(0.1,2.5){$\scriptstyle \sigma_{tot}/pb$}
\put(4,-.8){$ \scriptstyle \sqrt{s}/GeV$}
\put(7.1,2.5){$\scriptstyle \sigma_{tot}/pb$}
\put(11,-.8){$ \scriptstyle \sqrt{s}/GeV$}
\put(1.8,2.3){$ \scriptstyle (a)$}
\put(8.8,2.3){$ \scriptstyle (b)$}
\end{picture}
\end{center}
\vspace{2mm}
\caption{
        The total cross section $\sigma_{tot}$ 
        for $\mu^+\mu^- \rightarrow \tilde{\chi}^+_1 \tilde{\chi}^-_1$
        in (a) the mixed 
        and (b) the gaugino scenarios, for $m_{\tilde{\nu}_\mu} = 261$ 
	GeV, with no energy spread (continuous) and 
        with a finite energy resolution $R=0.06\% $ (dashed).}
\label{fig:Wien2}
\end{figure}

The cross section for chargino-pair production,
$\mu^+\mu^- \rightarrow \tilde{\chi}^+_1 \tilde{\chi}^-_1$,
around the $H$ and $A$ Higgs resonances with no energy spread and a
finite
energy resolution $R=0.06 \%$, is shown in Fig.~\ref{fig:Wien2}.
Since the energy separation between the resonances is in this case larger 
than their widths and the energy spread,
the two Higgs resonances can be separated clearly.
Since we assume here CP conservation,
the $H$ resonance is P-wave suppressed by the factor
$ (1-4 m^2_{\tilde{\chi}^\pm_1}/s) $.  
The peak of the $A$ resonance is thus higher than the $H$ peak
in both scenarios, although chargino couplings to $H$ are larger.
Comparing the scenarios,
the $A$ resonance is lower in the mixed scenario despite 
larger couplings and branching ratios, since also 
the $A$ decay width becomes larger, because of
decay channels into neutralinos.

The ratio of the Higgs-chargino couplings
\begin{equation}
  x:=\frac{c^2_{H\tilde{\chi}^+_1\tilde{\chi}^-_1}}
  {c^2_{A\tilde{\chi}^+_1\tilde{\chi}^-_1}}
\end{equation}
can be determined by measuring the ratio of the
cross sections on the Higgs resonance peaks:
\cite{ffp}
\begin{equation}
  r =  \frac{\sigma_{H}(m_H) + \sigma_{A}(m_H)   }
   {\sigma_{H}(m_A) + \sigma_{A}(m_A) }
      = \frac{\sigma_{tot}(m_H) - \sigma_{\gamma Z \tilde{\nu}}(m_H)   }
        {\sigma_{tot}(m_A) - \sigma_{\gamma Z \tilde{\nu}}(m_A) },
\end{equation}
where 
$\sigma_{H}(m_{H,A})$,
$\sigma_{A}(m_{H,A})$ and
$\sigma_{\gamma Z \tilde{\nu}}(m_{H,A})$
are the contributions
to the chargino-pair production cross section
from $H$ exchange, $A$ exchange and $\gamma/Z/\tilde{\nu}$ exchanges 
at the top of the $H,A$ resonances, respectively, 
neglecting the contribution of the lightest Higgs scalar $h$.
Interference between the two Higgs-boson exchange channels
vanishes when CP is conserved, as we assume here. It would be 
interesting to study the same reaction
in the MSSM with explicit loop-induced CP violation, along the lines
discussed in the previous section.
Interferences between the Higgs channels and the non-Higgs channels
are of order ${\cal O} (m_\mu/\sqrt{s})$,
and are therefore neglected.

\begin{figure}
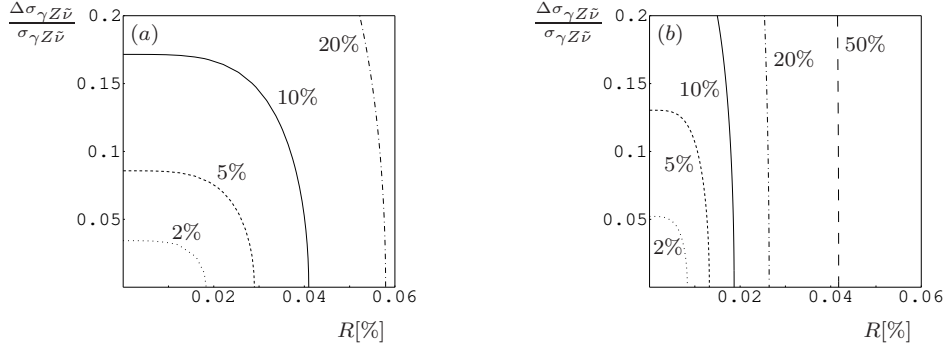

\begin{center}
\unitlength1cm
\begin{picture}(12,4)
\put(1,-0.5){\includegraphics{em.mps}}
\put(8,-0.5){\includegraphics{eg.mps}}
\put(1.75,3.35){$ \scriptstyle (a)$}
\put(8.75,3.35){$ \scriptstyle (b)$}
\put(0.1,3.5){$ \scriptstyle 
      \frac{\Delta\sigma_{\gamma Z \tilde{\nu}}}{
           \sigma_{\gamma Z \tilde{\nu}}}$}
\put(7.1,3.5){$ \scriptstyle 
      \frac{\Delta\sigma_{\gamma Z \tilde{\nu}}}{
           \sigma_{\gamma Z \tilde{\nu}}}$}
\put(4.5,-0.6){$ \scriptstyle R[\%]$}
\put(11.5,-0.6){$ \scriptstyle R[\%]$}
\put(2.3,0.7){$ \scriptstyle 2\%   $}
\put(2.9,1.5){$ \scriptstyle 5\%   $}
\put(3.7,2.5){$ \scriptstyle 10\%   $}   
\put(4.25,3.2){$ \scriptstyle 20\%   $}
\put(8.7,0.5){$ \scriptstyle 2\%   $} 
\put(8.85,1.6){$ \scriptstyle 5\%   $}
\put(9.05,2.6){$ \scriptstyle 10\%   $}     
\put(10.3,3.0){$ \scriptstyle 20\%   $}
\put(11.25,3.2){$ \scriptstyle 50\%   $}
\end{picture}
\end{center}
\caption{
        Relative error in the ratio of the Higgs-chargino couplings $x$
        as a function of the energy resolution and the relative error
        in the non-Higgs channels, 
        in (a) the mixed and (b) the gaugino scenarios. }
\label{fig:Wien3}
\end{figure}

We note that the ratio $r$, and therefore also $x$,
is independent of the chargino decay
characteristics. Then the  
error in 
the determination of the
Higgs-chargino couplings plotted in Fig.~\ref{fig:Wien3}
depends on the energy resolution 
of the muon beams
and on the error in 
the measurement of the non-Higgs contributions at the Higgs resonances in 
$r$, that can be estimated, e.g., from cross-section measurements
off the resonances. The effect of the energy resolution
on the cross sections and widths is larger if the widths are narrower.
With an energy resolution of $R=0.04\%$, 
the relative error on $x$ 
in the mixed scenario is larger than 10\%,
whereas in the gaugino scenario it lies around 50\%.
If, on the other hand, values of  $R\sim 0.01\%$ are achieved,
the error induced is in both cases of the order of 1\%.
With knowledge of the energy spread, the errors in the widths and
cross sections can be substantially
reduced~\cite{cern99}.
A detailed analysis of 
chargino and neutralino production at a \mm\ collider and
the precise determination of the Higgs couplings will be given 
in~\cite{ffp}.

We conclude that chargino production via $s$-channel Higgs exchange at a
\mm\ collider may allow a precise determination of the Higgs-chargino
couplings in the MSSM, if the resonances can be separated. We leave for a
future occasion discussions of the cases where the resonances overlap,
and when CP violation is important. 

\section{Conclusions}

We have seen in this chapter some of the physics opportunities offered by
\mm\ colliders operated as Higgs factories. Interest in these opportunities
has been stimulated by the possible existence of a light Higgs boson. 
The case outlined here depends indeed upon the mass of this lightest
Higgs boson, and, although all the indications are favourable, this 
remains an unknown parameter.

A first muon collider (FMC) operating around the peak of the Standard
Model (SM) Higgs boson, that is expected to weigh $\sim 120~\gevc$, as
seen in Fig.~\ref{fig:erler}, offers interesting prospects for precision
measurements of properties of the SM Higgs boson, such as its mass and
decay width. The mass could be measured to an unprecedented accuracy in
the sub-MeV region. A direct width determination would be possible to an
accuracy of ${\cal O}(1)$~MeV.
By varying judiciously the beam-energy spread, an interesting peak event
rate could be attained for a SM Higgs mass up to about $160~\gev$, and
measurements of decays into ${\bar b} b$, $\tau^+\tau^-$ and $W W^*$ may
be possible.
The accuracies obtained for these branching ratios are in the same ball
park as those expected at an $e^+e^-$ linear collider (LC).
However, they are highly dependent on the available luminosity and
the details of the detector, which are not yet fixed for the FMC.
Notice also that by combining with LC measurements the coupling of the 
Higgs boson to $\mu^+\mu^-$ could be determined with an accuracy of 
$\sim 4\%$ at the FMC~\cite{Barger:2001mi}. 

If supersymmetry plays a r\^ole at the electroweak scale, one expects
a richer Higgs sector, containing three neutral Higgs bosons $h,H,A$.
As we have shown, the production cross sections and branching ratios
of $h,H,A$ are very sensitive to supersymmetric radiative corrections.
As we have also shown, polarized \mm\ beams would offer in addition
interesting opportunities to explore CP violation in decay vertices and/or
Higgs-mass mixing.
The masses of $H,A$ may already be estimated quite accurately using FMC
measurements. A second Higgs factory (SMC) tuned to the twin $H, A$
peaks offers valuable prospects for measuring the two masses
independently (due to the fine energy resolution reachable at a 
\mm\ collider),
the couplings of the heavy Higgs bosons to SM and supersymmetric 
particles, as well as the opportunity for further interesting 
measurements of possible CP violation in the MSSM Higgs sector.

We have not discussed in this report the physics prospects offered by
a high-energy \mm\ collider. As is well known, this would have certain
advantages over a high-energy $e^+ e^-$ collider, notably in the beam
energy spread and in the accuracy with which the beam energy could be
calibrated using the precession of the $\mu^\pm$ polarization. However,
it is too early to know whether these advantages would be conclusive, and 
a multi-TeV \mm\ collider would presumably need to be preceded by one or
more lower-energy Higgs factories.

\section*{Acknowledgements}

S.K. thanks the CERN Theory Division for financial support 
during a stay at CERN. The work of D.G. is supported by the
European Commission TMR programme under the grant ERBFMBICT 983539.





\end{document}